\documentclass[iop,revtex4]{emulateapj}
\slugcomment{Re-submitted to ApJ: 5 May 2016}

\usepackage[backref,breaklinks,colorlinks,citecolor=blue]{hyperref}
\usepackage[all]{hypcap}
\usepackage{graphicx} 
\usepackage{mathrsfs} 
\usepackage{natbib}
\usepackage{amsmath}
\usepackage{float}
\usepackage{am
ssymb}
\usepackage{color}
\usepackage{captcont}

\newcommand{\as}{$^{\prime\prime}$}

\begin{document}

\title{ALMA Survey of Lupus Protoplanetary Disks I: Dust and Gas Masses}

\author{M. Ansdell$^{1}$, J. P. Williams$^{1}$, N. van der Marel$^{1,2}$, J. M. Carpenter$^{3}$, G. Guidi$^{6}$, M. Hogerheijde$^{2}$, G.S. Mathews$^{4}$, C.F. Manara$^{5}$, A. Miotello$^{2}$, A. Natta$^{6,7}$, I. Oliveira$^{8}$, M. Tazzari$^{9}$, L. Testi$^{9,6}$, E. F. van Dishoeck$^{2,10}$, S. E. van Terwisga$^{2}$}
 
\affil{$^1$Institute for Astronomy, University of Hawai`i at M\={a}noa, Honolulu, HI, USA}
\affil{$^2$Leiden Observatory, Leiden University, PO Box 9513, 2300 RA Leiden, The Netherlands}
\affil{$^3$ California Institute of Technology, 1200 East California Blvd, Pasadena, CA 91125, USA}
\affil{$^4$ Department of Physics and Astronomy, University of Hawaii, Honolulu, HI 96822, USA}
\affil{$^5$Scientific Support Office, Directorate of Science, European Space Research and Technology Centre (ESA/ESTEC), Keplerlaan 1, 2201 AZ Noordwijk, The Netherlands}
\affil{$^6$INAF-Osservatorio Astrofisico di Arcetri, Largo E. Fermi 5, I-50125 Firenze, Italy}
\affil{$^7$ Dublin Institute for Advanced Studies, School of Cosmic Physics, 31 Fitzwilliam Place, Dublin 2, Ireland}
\affil{$^8$ Observat\'{o}rio Nacional/MCTI, Rio de Janeiro, 20921-400, Brazil}
\affil{$^9$European Southern Observatory, Karl-Schwarzschild-Str. 2, D-85748 Garching bei M\"{u}nchen, Germany}
\affil{$^{10}$Max-Plank-Institut f\"{u}r Extraterrestrische Physik, Giessenbachstra\ss e 1, D-85748 Garching, Germany}

%========================= ABSTRACT =========================

\begin{abstract}

We present the first high-resolution sub-mm survey of both dust and gas for a large population of protoplanetary disks. Characterizing fundamental properties of protoplanetary disks on a statistical level is critical to understanding how disks evolve into the diverse exoplanet population. We use ALMA to survey 89 protoplanetary disks around stars with $M_{\ast}>0.1~M_{\odot}$ in the young (1--3~Myr), nearby (150--200~pc) Lupus complex. Our observations cover the 890~$\mu$m continuum and the $^{13}$CO and C$^{18}$O 3--2 lines. We use the sub-mm continuum to constrain $M_{\rm dust}$ to a few Martian masses (0.2--0.4~$M_{\oplus}$) and the CO isotopologue lines to constrain $M_{\rm gas}$ to roughly a Jupiter mass (assuming ISM-like $\rm {[CO]/[H_2]}$ abundance). Of 89 sources, we detect 62 in continuum, 36 in $^{13}$CO, and 11 in C$^{18}$O at $>3\sigma$ significance. Stacking individually undetected sources limits their average dust mass to $\lesssim6$ Lunar masses (0.03~$M_{\oplus}$), indicating rapid evolution once disk clearing begins. We find a positive correlation between $M_{\rm dust}$ and $M_{\ast}$, and present the first evidence for a positive correlation between $M_{\rm gas}$ and $M_{\ast}$, which may explain the dependence of giant planet frequency on host star mass. The mean dust mass in Lupus is 3$\times$ higher than in Upper Sco, while the dust mass distributions in Lupus and Taurus are statistically indistinguishable. Most detected disks have $M_{\rm gas}\lesssim1~M_{\rm Jup}$ and gas-to-dust ratios $<100$, assuming ISM-like $\rm {[CO]/[H_2]}$ abundance; unless CO is very depleted, the inferred gas depletion indicates that planet formation is well underway by a few Myr and may explain the unexpected prevalence of super-Earths in the exoplanet population.

\end{abstract}

\maketitle

%======================= INTRODUCTION ===================================

\section{INTRODUCTION}
\label{sec-intro}

The space-based {\it Kepler} transit survey \citep{2010Sci...327..977B} and long-term ground-based radial velocity surveys \citep{2010Sci...330..653H,2011arXiv1109.2497M} have opened the field of exoplanet statistics, revealing an unexpected diversity in exoplanet systems \citep{2015ARA&A..53..409W}. But how such diverse planetary systems form remains unclear, as similar demographic surveys of the preceding protoplanetary disks have been limited by the sensitivity and resolution of sub-mm arrays, which are our best tool for probing these cold and often faint objects \citep{2011ARA&A..49...67W}. Surveys of the optically thick infrared (IR) emission across young stellar clusters have constrained the disk dispersal timescale to $\sim$10~Myr, providing important checks on planet formation theories \citep{2001ApJ...553L.153H,2007ApJ...662.1067H,2009ApJS..181..321E}. However, surveys of the optically thin sub-mm emission are needed to probe the evolution of {\em bulk} dust and gas content; these fundamental properties dictate the planet-forming capacity of disks \cite[e.g.,][]{2005A&A...434..343A,2009A&A...501.1139M,2015A&A...575A..28B} and thus can help explain the diverse nature of the resulting exoplanet systems.

\bigskip
Population synthesis models are typically used to indirectly study how protoplanetary disks may evolve into planetary systems. Population synthesis models start from assumed initial disk properties (e.g., dust surface density) then use prescriptions for planet formation (e.g., core accretion) to explain observed exoplanet systems. What types of planetary systems form depend sensitively on the assumed disk properties \cite[e.g., higher-mass disks produce more massive planets;][]{2012A&A...541A..97M}, yet disk properties and their evolution have remained largely unconstrained for the average disk due to observational biases and limitations. For example, population synthesis models often assume the canonical interstellar medium (ISM) gas-to-dust ratio of $\sim$100 \citep{1978ApJ...224..132B}, yet observations of a small sample of Taurus disks suggest that this inherited value may decrease by a factor of $\sim$6 after just a few Myr \citep{2014ApJ...788...59W}.

A key factor limiting our understanding of disk evolution is the small number of protoplanetary disks with independently measured bulk dust and gas masses. Previous sub-mm surveys of young clusters concentrated on dust content \citep{2009ApJ...700.1502A,2011ApJ...736..135L,2013ApJ...771..129A,2013MNRAS.435.1671W,2014ApJ...787...42C,2015ApJ...806..221A}, but were often incomplete and {\em none} measured bulk gas masses due to observational constraints. Although a spectroscopic survey by \cite{2010A&A...510A..72F} showed that {\em inner} gas disk lifetimes are shorter than dust dissipation timescales, sensitive interferometers are needed to probe {\em bulk} gas content while avoiding confusion with cloud emission. Studying both gas and dust is critical because growing dust grains decouple from the gas and evolve differently, yet both components determine what types of planets may form. For example, super-Earths may result when giant planet cores form in gas-depleted disks, prohibiting the cores from rapidly accreting gaseous envelopes as predicted by core accretion theory \cite[e.g.,][]{1996Icar..124...62P,2004ApJ...604..388I}. If gas depletion is common in disks, this may help explain the unexpected prevalence of super-Earths and scarcity of gas giants seen in the exoplanet population \citep{2012ApJS..201...15H,2013ApJ...770...69P,2014PNAS..11112655M}.

\capstartfalse 
\begin{deluxetable}{lrrrr} 
\tabletypesize{\footnotesize} 
\centering 
\tablewidth{240pt} 
\tablecaption{Stellar Properties \label{tab-star}} 
\tablecolumns{5}  
\tablehead{ 
 \colhead{Source} 
&\colhead{$d$ (pc)} 
&\colhead{SpT} 
&\colhead{$M_{\ast}$/$M_{\odot}$} 
&\colhead{Ref} 
} 
\startdata 
Sz 65 & 150 & K7.0 & 0.76 $\pm$ 0.18 & 2 \\
Sz 66 & 150 & M3.0 & 0.31 $\pm$ 0.04 & 1 \\
J15430131-3409153 & 150 & ... & ... & ... \\
J15430227-3444059 & 150 & ... & ... & ... \\
J15445789-3423392 & 150 & M5.0 & 0.12 $\pm$ 0.03 & 1 \\
J15450634-3417378 & 150 & ... & ... & ... \\
J15450887-3417333 & 150 & M5.5 & 0.14 $\pm$ 0.03 & 2 \\
Sz 68 & 150 & K2.0 & 2.13 $\pm$ 0.34 & 2 \\
Sz 69 & 150 & M4.5 & 0.19 $\pm$ 0.03 & 1 \\
Sz 71 & 150 & M1.5 & 0.42 $\pm$ 0.11 & 1
\enddata 
\tablenotetext{}{References: (1) \cite{2014A&A...561A...2A}, (2) Alcal\'{a} et al. (in prep), (3) \cite{2013MNRAS.429.1001A}, (4) \cite{2011MNRAS.418.1194M}, (5) \cite{2008ApJS..177..551M}, (6) Cleeves et al. (in prep), (7) \cite{2015A&A...578A..23B}, (8) \cite{2008hsf2.book..295C}. Full table available online.} 
\end{deluxetable} 
\capstartfalse

Characterizing the evolution of protoplanetary disks on a statistical level, in both dust and gas, is therefore critical to understanding planet formation and the observed exoplanet population. The Atacama Large Millimeter/submillimeter Array (ALMA) now provides the sensitivity and resolution at sub-mm wavelengths to enable large-scale surveys of star-forming regions with ages spanning the observed disk dispersal timescale. In this work, we use ALMA to conduct the first high-resolution sub-mm survey, in both dust and gas, of a large population of protoplanetary disks. We study the young (1--3~Myr) and nearby (150--200~pc) Lupus star-forming region, as its proximity and youth make it an ideal target for a baseline study of early disk properties. 

We describe our sample selection in Section~\ref{sec-sample} and ALMA observations in Section~\ref{sec-observations}. The measured continuum and line fluxes are presented in Section~\ref{sec-results}, then converted to bulk dust and gas masses in Section~\ref{sec-analysis}. We identify correlations with stellar properties and examine disk evolution and planet formation in Section~\ref{sec-discussion}. Our findings are summarized in Section~\ref{sec-summary}.

 \capstartfalse
\begin{figure}[!ht]
\begin{centering}
\includegraphics[width=8.6cm]{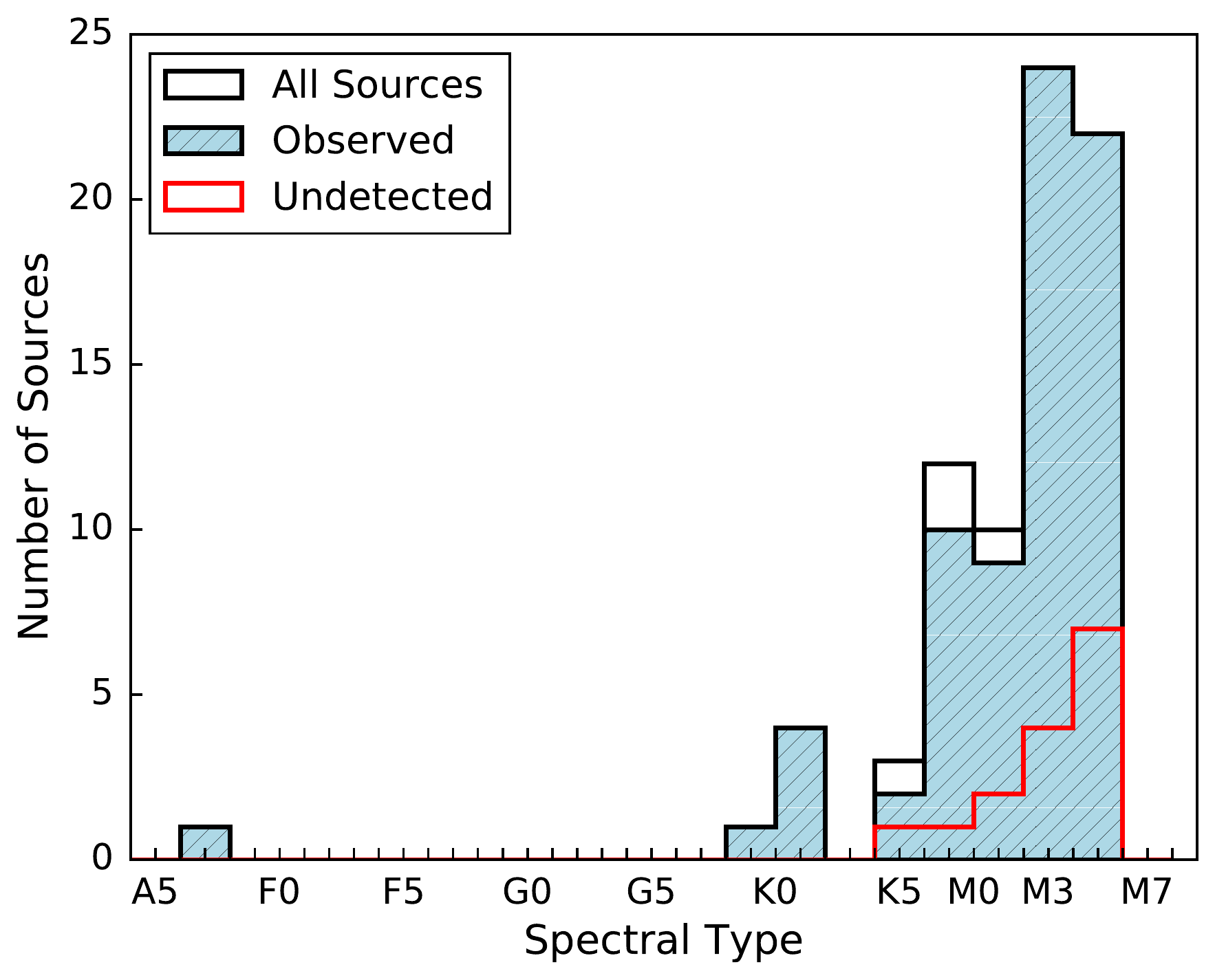}
\caption{\small Distribution of sources in our Lupus sample with known stellar spectral types (Table~\ref{tab-star}). The blue histogram shows sources with ALMA observations, the open histogram includes four sources for which we did not obtain ALMA observations (Section~\ref{sec-sample}), and the red histogram shows the locations of sources undetected in the continuum (Section~\ref{sec-cont}).}
\label{fig-spt}
\end{centering}
\end{figure}
\capstartfalse

 %======================= LUPUS SAMPLE  ==================================

\capstartfalse 
\begin{deluxetable*}{lccrccrrr} 
\tabletypesize{\footnotesize} 
\centering 
\tablewidth{500pt} 
\tablecaption{$890\mu$m Continuum Properties \label{tab-cont}} 
\tablecolumns{9}  
\tablehead{ 
 \colhead{Source} 
&\colhead{RA$_{\rm J2000}$} 
&\colhead{Dec$_{\rm J2000}$} 
&\colhead{$F_{\rm cont}$} 
&\colhead{rms} 
&\colhead{$a$} 
&\colhead{$i$} 
&\colhead{PA} 
&\colhead{$M_{\rm dust}$} \\ 
 \colhead{} 
&\colhead{} 
&\colhead{} 
&\colhead{(mJy)} 
&\colhead{(mJy beam$^{-1}$)} 
&\colhead{(arcsec)} 
&\colhead{(deg)} 
&\colhead{(deg)} 
&\colhead{($M_{\oplus}$)} 
} 
\startdata 
Sz 65 & 15:39:27.75 & -34:46:17.56 & 64.49 $\pm$ 0.32 & 0.30 & 0.171 $\pm$ 0.002 & 0 $\pm$ 0 & 0 $\pm$ 57 & 15.16 $\pm$ 0.08 \\
Sz 66 & 15:39:28.26 & -34:46:18.44 & 14.78 $\pm$ 0.29 & 0.27 & ... & ... & ... & 3.47 $\pm$ 0.07 \\
J15430131-3409153 & 15:43:01.29 & -34:09:15.40 & 0.01 $\pm$ 0.31 & 0.39 & ... & ... & ... & 0.00 $\pm$ 0.07 \\
J15430227-3444059 & 15:43:02.29 & -34:44:06.20 & 0.22 $\pm$ 0.27 & 0.34 & ... & ... & ... & 0.05 $\pm$ 0.06 \\
J15445789-3423392 & 15:44:57.90 & -34:23:39.50 & -0.05 $\pm$ 0.18 & 0.24 & ... & ... & ... & -0.01 $\pm$ 0.04 \\
J15450634-3417378 & 15:45:06.32 & -34:17:38.28 & 15.00 $\pm$ 0.40 & 0.34 & 0.096 $\pm$ 0.017 & 43 $\pm$ 28 & 24 $\pm$ 39 & 3.53 $\pm$ 0.09 \\
J15450887-3417333 & 15:45:08.85 & -34:17:33.81 & 46.27 $\pm$ 0.50 & 0.40 & 0.173 $\pm$ 0.005 & 45 $\pm$ 4 & -16 $\pm$ 5 & 10.87 $\pm$ 0.12 \\
Sz 68 & 15:45:12.84 & -34:17:30.98 & 150.37 $\pm$ 0.46 & 0.61 & 0.159 $\pm$ 0.002 & 34 $\pm$ 2 & -5 $\pm$ 3 & 35.34 $\pm$ 0.11 \\
Sz 69 & 15:45:17.39 & -34:18:28.66 & 16.96 $\pm$ 0.28 & 0.24 & 0.092 $\pm$ 0.012 & 69 $\pm$ 21 & -39 $\pm$ 11 & 3.99 $\pm$ 0.07 \\
Sz 71 & 15:46:44.71 & -34:30:36.05 & 166.04 $\pm$ 0.63 & 0.37 & 0.558 $\pm$ 0.003 & 40 $\pm$ 0 & 42 $\pm$ 1 & 39.02 $\pm$ 0.15
\enddata 
\tablenotetext{}{Full table available online.} 
\end{deluxetable*} 
\capstartfalse

\capstartfalse
\begin{figure*}[!ht]
\begin{centering}
\includegraphics[width=18cm]{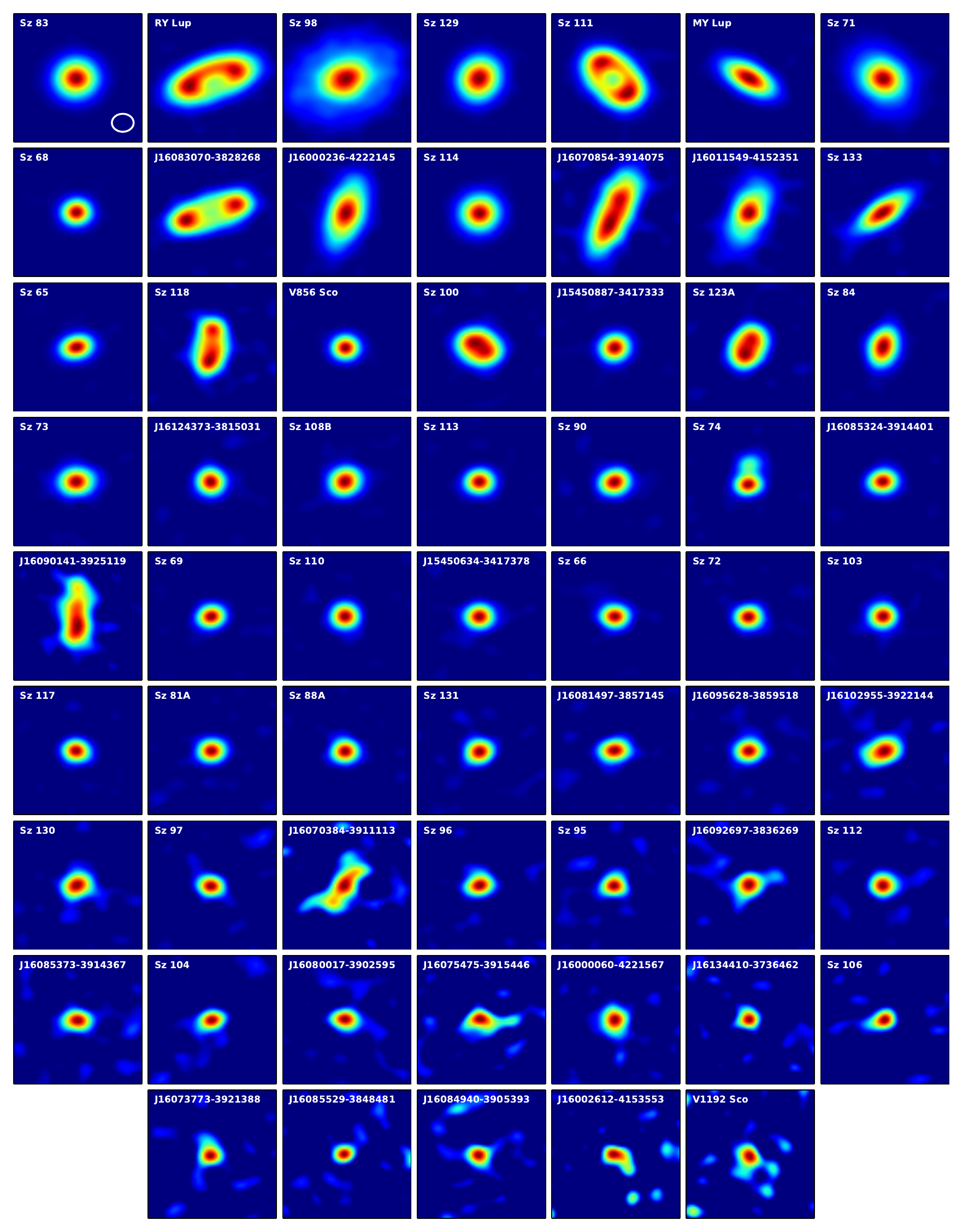}
\caption{\small 890~$\mu$m continuum images of the 61 Lupus disks detected in our ALMA Cycle 2 program (this excludes Sz~82, which was observed by Cleeves et al., in prep), ordered by decreasing continuum flux density (as reported in Table~\ref{tab-cont}). Images are 2\as{}$\times$2\as{} and the typical beam size is shown in the first panel.}
\label{fig-gallery}
\end{centering}
\end{figure*}
\capstartfalse

\section{LUPUS SAMPLE \& COMPLETENESS}
\label{sec-sample}

The Lupus complex consists of four main star-forming clouds (Lupus I--IV) and is one of the youngest and closest star-forming regions \cite[see review in][]{2008hsf2.book..295C}. In general, Lupus III is at $\sim$200 pc while Lupus I, II, and IV are at $\sim$150 pc, though distance estimates vary due to the depth of the complex. The age of Lupus is $\sim$1--2 Myr \cite[][and references therein]{2008hsf2.book..295C} but may be as old as $3\pm2$ Myr \citep{2014A&A...561A...2A}. Lupus I, III, and IV were observed for the c2d {\it Spitzer} legacy project \citep{2009ApJS..181..321E} and shown to have high disk fractions \cite[70--80\%;][]{2008ApJS..177..551M} indicating young disk populations. Lupus II is also an active star-forming region \citep{1977ApJS...35..161S} and contains RU Lup, one of the most active T Tauri stars known.

Our sample consists of Young Stellar Objects (YSOs) in Lupus  I--IV that are more massive than brown dwarfs (i.e., $M_{\ast}>0.1~M_{\odot}$) and host protoplanetary disks (i.e., have Class II or Flat IR spectra). We compiled our sample from sources fitting these criteria in the Lupus disk catalogues available at the time of our ALMA Cycle~2 proposal \citep{2008ApJS..177..551M,2008hsf2.book..295C,2011MNRAS.418.1194M,1994AJ....108.1071H}. Disk classifications were taken from the literature and primarily derived from the IR spectral index slope ($\alpha$) between the 2MASS $K_{\rm S}$ (2~$\mu$m) and {\it Spitzer} MIPS-1 (24~$\mu$m) bands; for sources without {\it Spitzer} data, disk classifications were approximated from their IR excesses and/or accretion signatures (e.g., H$\alpha$). The preliminary stellar masses used for our sample selection were estimated by fitting absolute $J$-band magnitudes to the 3~Myr \cite{2000A&A...358..593S} model isochrone.

We checked this sample against more recently published Lupus disk catalogues, which utilized updated {\it Spitzer} \citep{2015ApJS..220...11D} and {\it Herschel} \citep{2015A&A...578A..23B} data. These updated catalogues included one additional source that fit our selection criteria, V1094 Sco (RXJ1608.6-3922), which we added to our sample but did not observe in our ALMA Cycle 2 program. We also removed eight sources from our sample whose follow-up VLT/X-Shooter spectra (Alcal\'{a} et al., in prep) revealed as background giants due to discrepant surface gravities and radial velocities (Frasca et al., in prep.); these sources were observed (but undetected) during our ALMA Cycle 2 program (see Table~\ref{tab-rejected}). 

Table~\ref{tab-star} gives the 93 Lupus disks that fit our selection criteria. Our adopted nomenclature uses primarily Sz or 2MASS names, supplemented with c2d names. Our ALMA Cycle 2 program did not observe Sz~76, Sz~77, Sz~82, Sz~91, and V1094~Sco. However, for Sz~82 (IM~Lup) we utilize existing ALMA data (Cleeves et al., in prep) in this work. Note that although RXJ1556.1-3655 and RXJ1615.3-3255 are listed as Class~II Lupus disks in the literature \citep{2010ApJ...724..835W,2011ApJ...732...42A}, we excluded them from our sample as their coordinates are off the Lupus I-IV clouds \cite[e.g.,][]{1999A&A...345..965C}. Thus we obtained ALMA data for 89 (out of 93) sources in our sample for a 96\% completeness rate.

Table~\ref{tab-star} provides some basic stellar properties for our Lupus sample. Stellar masses ($M_{\ast}$) were derived for the 69 sources with spectroscopically determined stellar effective temperatures ($T_{\rm eff}$) and luminosities ($L_{\ast}$) using \cite{2000A&A...358..593S} evolutionary models; the median $M_{\ast}$ uncertainty is $\sim$16\%. For details on the derivation of these $M_{\ast}$ values, as well as their associated $T_{\rm eff}$ and $L_{\ast}$ values, see \cite{2014A&A...561A...2A} and Alcal\'{a} et al. (in prep). The remaining 20 sources are highly obscured, making it difficult to derive accurate stellar properties, thus we do not provide $M_{\ast}$ values for these sources in Table~\ref{tab-star}.

\capstartfalse 
\begin{deluxetable*}{lrrrrrrr} 
\tabletypesize{\footnotesize} 
\centering 
\tablewidth{500pt} 
\tablecaption{Gas Properties \label{tab-gas}} 
\tablecolumns{8}  
\tablehead{ 
 \colhead{Source} 
&\colhead{$F_{\rm 13CO}$} 
&\colhead{$E_{\rm 13CO}$} 
&\colhead{$F_{\rm C18O}$} 
&\colhead{$E_{\rm C18O}$} 
&\colhead{$M_{\rm gas}$} 
&\colhead{$M_{\rm gas,min}$} 
&\colhead{$M_{\rm gas,max}$} \\ 
 \colhead{} 
&\colhead{(mJy~km~s$^{-1}$)} 
&\colhead{(mJy~km~s$^{-1}$)} 
&\colhead{(mJy~km~s$^{-1}$)} 
&\colhead{(mJy~km~s$^{-1}$)} 
&\colhead{($M_{\rm Jup}$)} 
&\colhead{($M_{\rm Jup}$)} 
&\colhead{($M_{\rm Jup}$)} 
} 
\startdata 
Sz 65 & 971 & 128 & 415 & 105 & 0.7 & 0.3 & 10.5 \\
Sz 66 & 153 & 45 & $<111$ & ... & 0.2 & ... & 1.0 \\
J15430131-3409153 & $<162$ & ... & $<192$ & ... & $<1.0$ & ... & ... \\
J15430227-3444059 & $<138$ & ... & $<171$ & ... & $<1.0$ & ... & ... \\
J15445789-3423392 & $<84$ & ... & $<102$ & ... & $<0.3$ & ... & ... \\
J15450634-3417378 & 356 & 111 & $<174$ & ... & 0.1 & ... & 3.1 \\
J15450887-3417333 & 759 & 87 & 573 & 145 & 3.2 & 1.0 & 10.5 \\
Sz 68 & 915 & 133 & 444 & 132 & 0.8 & 0.3 & 10.5 \\
Sz 69 & 466 & 74 & $<102$ & ... & 0.2 & ... & 3.1 \\
Sz 71 & 1298 & 107 & $<111$ & ... & 0.3 & ... & 1.0
\enddata 
\tablenotetext{}{Full table available online.} 
\end{deluxetable*} 
\capstartfalse

 %======================= ALMA OBSERVATIONS  =============================

\section{ALMA OBSERVATIONS}
 \label{sec-observations}
 
Our ALMA Cycle 2 observations (Project ID: 2013.1.00220.S) were obtained on 2015 June 14 (AGK-type sources and unknown spectral types) and  2015 June 15 (M-type sources). The continuum spectral windows were centered on  328.3, 340.0, and 341.8~GHz with bandwidths of 1.875, 0.938, and 1.875~GHz and channel widths of 15.625, 0.244, and 0.977~MHz, respectively. The bandwidth-weighted mean continuum frequency was 335.8~GHz (890~$\mu$m). The spectral setup included two windows covering the $^{13}$CO and C$^{18}$O 3--2 transitions; these spectral windows were centered on 330.6 and 329.3~GHz, respectively, with bandwidths of 58.594~MHz, channel widths of 0.122~MHz, and velocity resolutions of 0.11~km~s$^{-1}$. 

The array configuration used 37 12-m antennas for the M-type sample and 41 12-m antennas for the AGK-type sample with baselines of 21.4--783.5~m. We integrated for 30 sec per source on the AGK-type sample and 1 min per source on the M-type sample for an average rms of 0.41 and 0.25 mJy beam$^{-1}$, respectively. Data calibration and imaging were performed using CASA 4.4.0. The data were pipeline calibrated by NRAO and included flux, phase, bandpass, and gain calibrations. Flux calibration used observations of Titan, passband calibration used observations of J1427-4206, and gain calibration used observations of J1604-4228 or J1610-3958. We estimated an absolute flux calibration error of 10\% based on variations in the gain calibrators. 

We extracted continuum images from the calibrated visibilities by averaging over the continuum channels and cleaning with a Briggs robust weighting parameter of $+0.5$ for unresolved sources and $-1.0$ for resolved sources. The average continuum beam size was 0.34\as{}$\times$0.28\as{} ($\sim$50$\times$40~AU at 150~pc). We extracted $^{13}$CO and C$^{18}$O 3--2 channel maps from the calibrated visibilities by subtracting the continuum and cleaning with a Briggs robust weighting parameter of $+0.5$ in all cases due to the weakness of the line emission. Zero-moment maps were created by integrating over the velocity channels showing line emission; the appropriate velocity range was determined for each source by visual inspection of the channel map and spectrum. If no emission was visible, we summed across the average velocity range of the detected sources ($\pm2.3$~km~s$^{-1}$) from the average radial velocity of Lupus I--IV sources \cite[$3.7$~km~s$^{-1}$;][]{2013A&A...558A..77G}. Note that the dispersion around this average radial velocity value is small \cite[$\pm0.4$~km~s$^{-1}$;][]{2013A&A...558A..77G}.

 %======================= ALMA RESULTS  ==================================

\section{ALMA RESULTS}
\label{sec-results}

\subsection{Continuum Emission  \label{sec-cont}}

We measured continuum flux densities by fitting elliptical Gaussians to the visibility data with {\it uvmodelfit} in CASA. The elliptical Gaussian model has six free parameters: integrated flux density ($F_{\rm cont}$), FWHM along the major axis ($a$), aspect ratio of the axes ($r$), position angle (PA), right ascension offset from the phase center ($\Delta\alpha$), and declination offset from the phase center ($\Delta\delta$). If the ratio of $a$ to its uncertainty was less than five, a point-source model with three free parameters ($F_{\rm cont}$, $\Delta\alpha$, $\Delta\delta$) was fitted to the visibility data instead. 

For disks with resolved structure, flux densities were measured from continuum images using circular aperture photometry. The aperture radius for each source was determined by a curve-of-growth method, in which we applied successively larger apertures until the measured flux density leveled off. Uncertainties were estimated by taking the standard deviation of the flux densities measured within a similarly sized aperture placed randomly within the field of view but away from the source.

Table~\ref{tab-cont} gives our measured 890~$\mu$m continuum flux densities and associated uncertainties. The uncertainties are statistical errors and do not include the 10\% absolute flux calibration error (Section~\ref{sec-observations}). For Sz~82, we report the ALMA continuum measurement from Cleeves et al. (in prep). Of the 89 sources, 62 were detected with $>3\sigma$ significance; the continuum images of the 61 sources detected by our ALMA Cycle 2 program (i.e., excluding Sz~82) are shown in Figure~\ref{fig-gallery}. Images for all sources observed by our ALMA Cycle 2 program are shown in Figure~\ref{fig-zoo}. In Table~\ref{tab-cont} we also provide the fitted source centers output by {\it uvmodelfit} for detections, or the phase centers of our ALMA observations for non-detections. For sources fitted with an elliptical Gaussian model, we give $a$ and PA as well as the inclination, $i$, derived from $r$ assuming circular disk structure. We also list the image rms, derived from a 4--9\as{} radius annulus centered on the fitted or expected source position for detections or non-detections, respectively.

\subsection{Line Emission \label{sec-line}}

We measured $^{13}$CO and C$^{18}$O 3--2 integrated flux densities and associated uncertainties from our ALMA zero-moment maps (Section~\ref{sec-observations}) using the same aperture photometry method described above for structured continuum sources (Section~\ref{sec-cont}). For non-detections, we took an upper limit of $3\times$ the uncertainty when using an aperture of the same size as the beam ($\sim$0.3\as{}). Table~\ref{tab-gas} gives our measured integrated flux densities or upper limits. For Sz~82, we report ALMA line measurements from Cleeves et al. (in prep). Of the 89 targets, 36 were detected in $^{13}$CO while only 11 were detected in C$^{18}$O with $>3\sigma$ significance. All sources detected in C$^{18}$O were also detected in $^{13}$CO, and all sources detected in $^{13}$CO were also detected in the continuum.

%============================ ANALYSIS ===================================
     
\section{PROPERTIES OF LUPUS DISKS}
\label{sec-analysis}
 
 \subsection{Dust Masses \label{sec-dust}}
 
Assuming dust emission at sub-mm wavelengths is isothermal and optically thin, the sub-mm continuum flux at a given wavelength ($F_{\nu}$) is directly related to the mass of the emitting dust ($M_{\rm dust}$), as shown in \cite{1983QJRAS..24..267H}:

\begin{equation}
M_{\rm dust}=\frac{F_{\nu}d^{2}}{\kappa_{\nu}B_{\nu}(T_{\rm dust})} \approx 7.06\times10^{-7} \left(\frac{d}{150}\right)^{2}F_{890\mu {\rm m}},
\label{eqn-mass}
\end{equation}

where $B_{\nu}(T_{\rm dust})$ is the Planck function for a characteristic dust temperature of $T_{\rm dust}=20$~K, the median for Taurus disks \citep{2005ApJ...631.1134A}. We took the dust grain opacity, $\kappa_{\nu}$, as 10 cm$^{2}$ g$^{-1}$ at 1000 GHz and used an opacity power-law index of $\beta=1$ \citep{1990AJ.....99..924B}. We used distances, $d$, from Table~\ref{tab-star}. 

Table~\ref{tab-cont} gives the $M_{\rm dust}$ values for our Lupus sample, derived from the 890 $\mu$m continuum flux densities and associated uncertainties measured in Section~\ref{sec-cont}; the top panel of Figure~\ref{fig-g2d} shows $M_{\rm dust}$ for all the continuum-detected disks. We employed this simple approach, which estimates disk-averaged dust masses assuming a single grain opacity and disk temperature, to ease interpretation and comparison with surveys of other regions (Section~\ref{sec-compare}). In particular, we do not scale $T_{\rm dust}$ with $L_{\ast}$ \citep{2013ApJ...771..129A} because: we do not have $L_{\ast}$ values for every star in our sample (Secton~\ref{sec-sample}); different surveys apply different methods of deriving stellar parameters; and our sample is dominated by M-type stars, which limits the potential affect on our inferred dust masses.

 \subsection{Gas Masses \label{sec-gas}}

\capstartfalse
\begin{figure*}[!ht]
\begin{centering}
\includegraphics[width=18.cm]{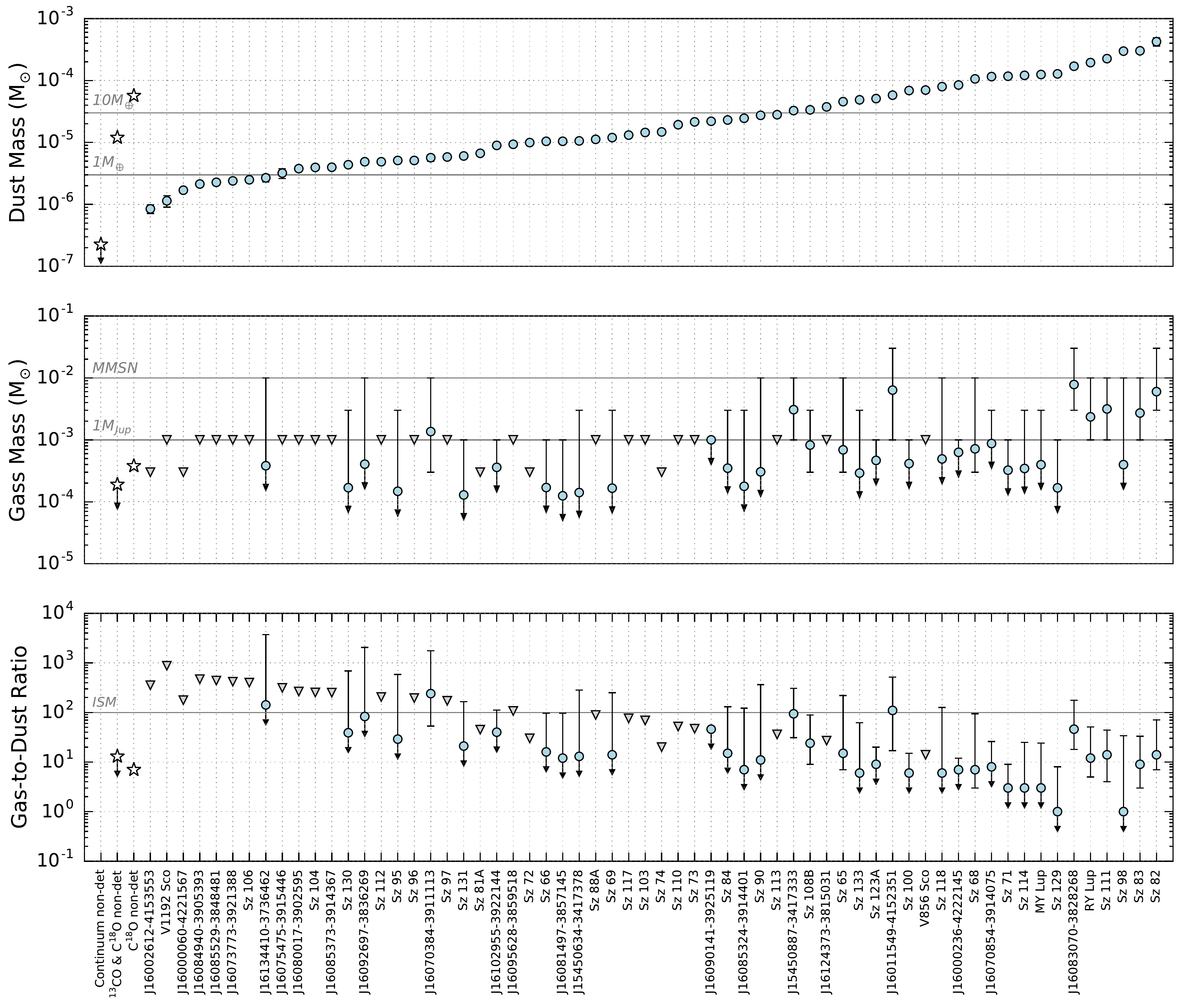}
\caption{\small Dust masses (top), gas masses (middle), and gas-to-dust ratios (bottom) for continuum-detected sources in our Lupus sample. Blue points indicate detections and gray triangles indicate upper limits. Dust masses are from Table~\ref{tab-cont} and described in Section~\ref{sec-dust}; the associated error bars include the 10\% absolute flux calibration uncertainty. Gas masses and associated ranges are from Table~\ref{tab-gas} and described in Section~\ref{sec-gas}; error bars with downward arrows indicate sources detected in $^{13}$CO but not C$^{18}$O, for which we did not place lower limits on their gas masses. Gas-to-dust ratios and associated ranges are directly calculated from the dust masses and the range of possible gas masses. Stars show the results of our stacking analysis (Section~\ref{sec-stacks}).}
\label{fig-g2d}
\end{centering}
\end{figure*}
\capstartfalse

Measuring the gas content in circumstellar disks is essential for a complete understanding of planet formation. The CO line observations in this work provide a means to estimate bulk gas masses independently from the dust. Physical-chemical models of protoplanetary disks have shown that a significant fraction of CO is sufficiently warm to avoid freeze-out and also sufficiently shielded from UV radiation to avoid photodissociation \citep{2002A&A...386..622A}. This structure has been simulated by \cite{2014ApJ...788...59W} (hereafter WB14) in parametrized models to show that the surviving CO gas fraction is generally large, except for exceptionally cold or low-mass disks. In particular, combining the $^{13}$CO and C$^{18}$O isotopologue lines, with their moderate-to-low optical depths, provides a relatively simple and robust proxy of bulk gas content in protoplanetary disks.

\capstartfalse
\begin{figure*}
\begin{centering}
\includegraphics[width=18.cm]{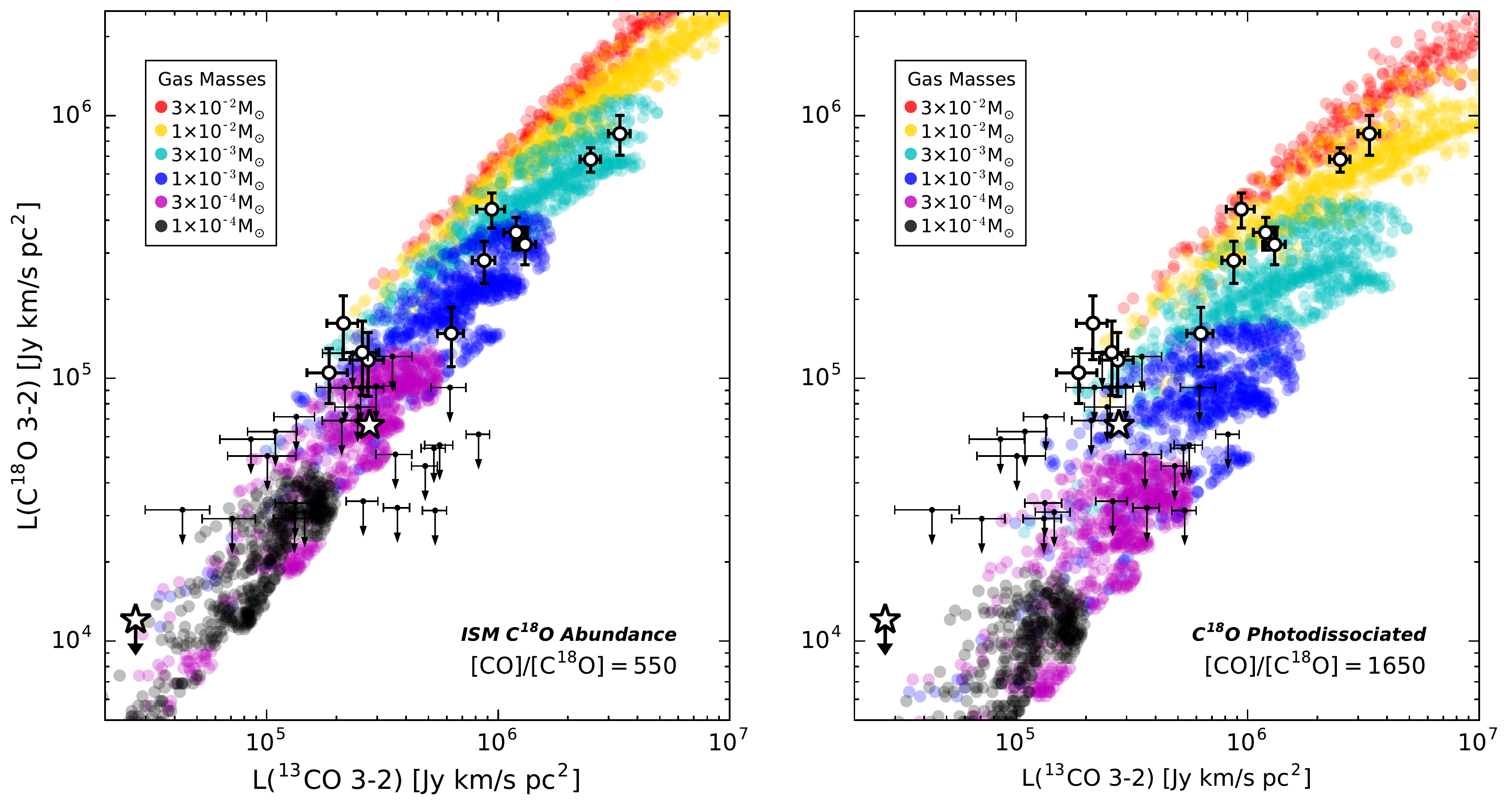}
\caption{\small $^{13}$CO and C$^{18}$O 3--2 line luminosities for determining gas masses (Section~\ref{sec-gas}). Colored points show the WB14 model grids color-coded by gas mass. The two panels show different values for the [C$^{18}$O]/[CO] isotopologue ratio, the ISM value of 550 (left) and 3$\times$ reduced abundance (right), to account for isotope-selective photodissociation. The 11 Lupus disks with both lines detected are plotted as white circles, and the 25 Lupus disks with only $^{13}$CO detections are plotted as black circles with arrows indicating 3$\sigma$ upper limits on C$^{18}$O. Error bars include both the statistical uncertainties (Table~\ref{tab-gas}) and a 10\% flux calibration error. Stars show locations of the stacked non-detections (Section~\ref{sec-stacks}); error bars are roughly the size of the symbol.}
\label{fig-gas}
\end{centering}
\end{figure*}
\capstartfalse

We compare our measured $^{13}$CO and C$^{18}$O 3--2 line luminosities to the WB14 model grids in Figure~\ref{fig-gas}. The two panels show different values for the C$^{18}$O isotopologue abundance: the ISM value (left) and a factor of 3 lower (right). The reduced C$^{18}$O abundance was required to fit some of the Taurus disk observations in WB14, and is similarly necessary to fit some of the Lupus disk observations in this work; specifically, the left panel of Figure~\ref{fig-gas} shows that some upper limits do not match any model grid points when assuming an ISM-like C$^{18}$O abundance. The physical reasoning for a reduced C$^{18}$O abundance is CO isotope-selective photodissociation \citep{1988ApJ...334..771V}, which has been modeled in detail for protoplanetary disks in \cite{2014A&A...572A..96M}. The empirical factor of 3 used in Figure~\ref{fig-gas} is sufficient to fit our Lupus observations and lies within the range of models in \cite{2014A&A...572A..96M} for massive disks. Although our observed fluxes do not match the models of \cite{2014A&A...572A..96M} for low-mass disks, those models covered a limited set of disk parameters and will be expanded to a larger model grid to interpret the CO isotopologue detections in Lupus with more sophisticated treatment of isotope-selective effects (Miotello et al., submitted).

Our derived gas masses are given in Table~\ref{tab-gas}. We determined these gas masses by comparing our $^{13}$CO and C$^{18}$O line luminosity measurements or upper limits to the WB14 model grids. We considered both WB14 model grids (ISM and 3$\times$ reduced C$^{18}$O abundance) in order to take into account possible isotope-selective photodissociation effects. The line luminosity uncertainties included the statistical errors in Table~\ref{tab-gas} and a 10\% absolute flux calibration error (added in quadrature). For the 11 sources detected in both $^{13}$CO and C$^{18}$O, we calculated the mean (in log space) of the WB14 model grid points within $\pm$3$\sigma$ of our measured $^{13}$CO and C$^{18}$O line luminosities ($M_{\rm gas}$), and also set upper ($M_{\rm gas,max}$) and lower ($M_{\rm gas,min}$) limits based on the maximum and minimum WB14 model grid points consistent with the data. For the 25 sources with $^{13}$CO detections and C$^{18}$O upper limits, we similarly calculated $M_{\rm gas}$ and $M_{\rm gas,max}$ but set $M_{\rm gas,min}$ to zero as the effect of isotope-selective photodissociation may be stronger for low-mass disks (Miotello et al., submitted). For the 53 disks undetected in both lines, we set only upper limits to the gas masses using the maximum model grid points consistent with the $^{13}$CO and C$^{18}$O line luminosity upper limits.

Figure~\ref{fig-g2d} (middle panel) shows the derived gas masses or upper limits for the continuum-detected sources in our Lupus sample. The disk gas masses are very low, in most cases much less than the Minimum Mass Solar Nebula (MMSN). For disks with at least one line detection, the inferred gas-to-dust ratios are almost universally below the ISM value of 100 (Figure~\ref{fig-g2d}, bottom panel). The gas-to-dust ratio upper limits are poor for low-mass disks with weak emission, but are more stringent for sufficiently massive disks ($M_{\rm dust} > 10~M_\oplus$). These low gas masses and low gas-to-dust ratios confirm the findings of WB14, who presented a similar result for a small sample of nine Class II disks around K/M-type sources in Taurus. We discuss the implications and caveats of these findings in Section~\ref{sec-lowgas}.

\subsection{Stacked Non-detections \label{sec-stacks}}

\capstartfalse
\begin{figure*}
\begin{centering}
\includegraphics[width=17.5cm]{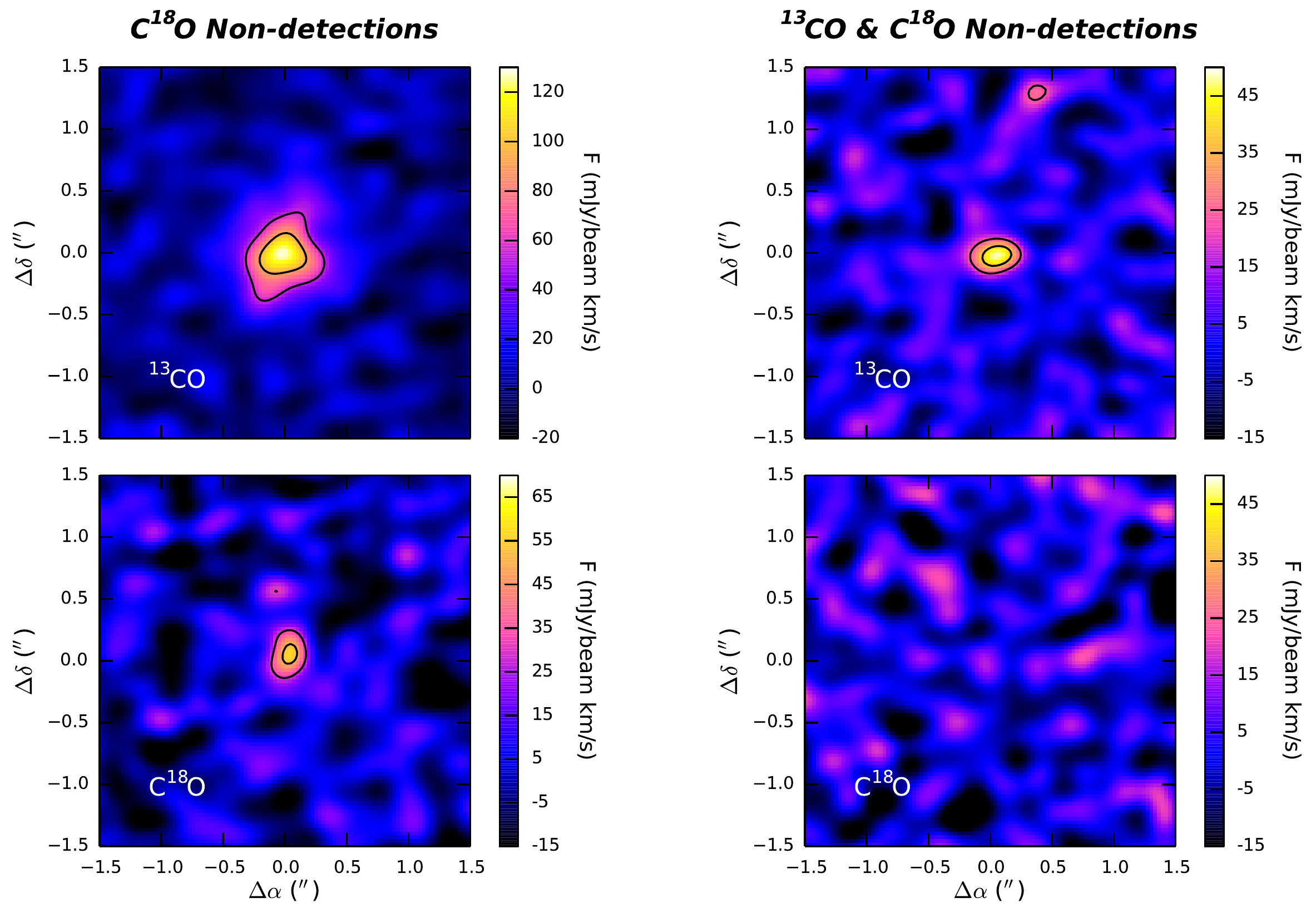}
\caption{\small Stacks of individually undetected sources (Section~\ref{sec-stacks}). The left panel shows stacks of sources detected in the continuum and $^{13}$CO, but not C$^{18}$O. The right panel shows stacks of sources detected in the continuum, but neither $^{13}$CO nor C$^{18}$O. The top panels show $^{13}$CO stacks and the bottom panels show C$^{18}$O stacks. Contour lines are 3$\sigma$ and 5$\sigma$ levels.}
\label{fig-stack}
\end{centering}
\end{figure*}
\capstartfalse

We performed a stacking analysis to constrain the average dust and gas mass of individually undetected sources. Before stacking, we centered each image on the expected source location and scaled the flux to 200 pc. We then measured flux densities in the stacked images using aperture photometry as in Section~\ref{sec-cont}. We confirmed that the source locations were known to sufficient accuracy for stacking by measuring the average offset of the detected sources from their phase centers: we found $\langle\Delta\alpha\rangle=-0.15$\as{} and $\langle\Delta\delta\rangle=-0.22$\as{}, both smaller than the average beam size. Moreover, because the dispersion around the mean radial velocity of Lupus I-IV sources is much smaller than the velocity range over which we integrated the zero-moment maps (Section~\ref{sec-observations}), any radial velocity differences among the gas non-detections should have negligible effects on the stacking.

We first stacked the 27 continuum non-detections, but did not find a significant mean signal in the continuum, $^{13}$CO, or C$^{18}$O stacks. The lack of line emission is expected given the undetected continuum, but the absence of continuum emission is surprising given the sensitivity of the stacked image. We measured a mean signal of $0.08\pm0.06$ mJy, which gives a 3$\sigma$ upper limit on the average dust mass of individually undetected continuum sources of $\sim$6 Lunar masses (0.03~$M_{\oplus}$), comparable to debris disk levels \citep{2008ARA&A..46..339W}. The stark contrast between the detected and undetected continuum sources (see Figure~\ref{fig-g2d}, top panel) suggests that protoplanetary disks evolve quickly to debris disk levels once disk clearing begins \citep{2014prpl.conf..475A}.

We then stacked the 25 sources detected in the continuum and $^{13}$CO, but not C$^{18}$O. We measured a continuum mean signal of $45.25\pm0.20$ mJy and a $^{13}$CO mean signal of $586\pm27$ mJy~km~s$^{-1}$ (Figure~\ref{fig-stack}, upper left panel). Interestingly, the stacking also revealed a significant mean signal for C$^{18}$O of $132\pm20$ mJy~km~s$^{-1}$ (Figure~\ref{fig-stack}, lower left panel). The stacked continuum flux corresponds to $M_{\rm dust}\sim19~M_{\oplus}$ and the stacked line fluxes correspond to $M_{\rm gas}\sim0.4~M_{\rm Jup}$ (Figure~\ref{fig-gas}), giving an average gas-to-dust ratio of only $\sim$7 for sources detected in the continuum and $^{13}$CO, but not C$^{18}$O (Figure~\ref{fig-g2d}, bottom panel).

Finally, we stacked the 26 sources detected in the continuum, but undetected in both $^{13}$CO and C$^{18}$O. We measured a continuum mean signal of $9.53\pm0.13$~mJy.  The stacking revealed a significant mean signal for $^{13}$CO (Figure~\ref{fig-stack}, upper right panel), but not C$^{18}$O (Figure~\ref{fig-stack}, lower right panel); the stacked gas fluxes were $54\pm7$~mJy~km~s$^{-1}$ and $3\pm8$~mJy~km~s$^{-1}$, respectively. The continuum flux corresponds to $M_{\rm dust}\sim4~M_{\oplus}$ while the $^{13}$CO line flux and C$^{18}$O upper limit correspond to $M_{\rm gas}\lesssim0.2~M_{\rm Jup}$ (Figure~\ref{fig-gas}), for an average gas-to-dust ratio of $\lesssim$13 for disks detected in the continuum but undetected in $^{13}$CO and C$^{18}$O (Figure~\ref{fig-g2d}, bottom panel).

 \subsection{Transition Disks \label{sec-TD}}

Transition disks (TDs) are protoplanetary disks with inner cavities and/or annular gaps in their dust distributions. TDs were initially identified by the mid-IR deficits in their Spectral Energy Distributions (SEDs), indicating a lack of warm micron-sized dust grains close to the central star \cite[see review in][]{2014prpl.conf..497E}. Inner disk clearings and dust rings were later confirmed with resolved mm continuum images \cite[e.g.,][]{2011ApJ...732...42A}. TDs are now considered to be sites of ongoing disk evolution, and their dust and gas distributions may be in some cases signposts of embedded planets clearing gaps in the disk \cite[e.g.,][]{2012A&A...545A..81P,2014ApJ...783L..13P,2015A&A...579A.106V,2016A&A...585A..58V,2015ApJ...805...21C}

Our ALMA continuum images show several Lupus disks with signatures of inner dust cavities. Three disks (2MASS~J16083070-3828268, RY~Lup, Sz~111) show clearly resolved dust rings with cavity diameters of $\sim$0.8\as{} ($\sim$80~AU radius at 200~pc). 2MASS~J16083070-3828268 and Sz~111 were previously recognized as TDs from the mid-IR deficits in their SEDs \citep{2008ApJS..177..551M}. Although RY~Lup was not previously identified as a TD by its SED \citep{2009A&A...499..137M}, this is likely because its strong 10~$\mu$m silicate emission feature, seen in its IRS spectrum \citep{2006ApJ...639..275K}, washes out the mid-IR dip in its broadband fluxes. 

Three sources (Sz~100, Sz~123, 2MASS~J16070854-3914075) show possible cavities with diameters of $\lesssim$0.4\as{}. Sz~123A is a known TD candidate from the IR deficit in its SED \citep{2008ApJS..177..551M,2015A&A...578A..23B}, though its mid-IR flux is potentially confused by its nearby companion Sz~123B, which is undetected in our ALMA images. The SED of Sz~100 is consistent with a primordial disk \citep{2008ApJS..177..551M}, though the broadband continuum is also likely affected by the bright silicate features seen in its IRS spectrum (Oliveira et al., in prep.). 2MASS~J16070854-3914075 shows small excesses at all IR wavelengths in its SED.

Six sources in our sample were identified as TD candidates in the literature, but do not exhibit cavities in our ALMA images: Sz~84 \citep{2010ApJ...718.1200M}, MY~Lup \citep{2012ApJ...749...79R}, Sz~112, 2MASS~J16011549-4152351 \citep{2016arXiv160307255V}, 2MASS~J16102955-3922144, and 2MASS~J16081497-3857145 \citep{2015A&A...578A..23B}. This implies that the dust cavities of these disks, if present, must be $\lesssim$0.3\as{} in diameter ($\lesssim$25~AU radius at 150~pc). We also did not observe two previously identified TD candidates in Lupus (Section~\ref{sec-sample}): Sz~76 \citep{2016arXiv160307255V} and Sz~91 \citep{2014ApJ...783...90T,2015A&A...578A..23B,2015ApJ...805...21C}. 

Thus the fraction of TDs in our Lupus sample with resolved cavities is 10\% (6/62) when considering only continuum detections. This increases to 19\% (12/62) when including the known TDs in our sample that were unresolved in our ALMA continuum images. This is consistent with previous estimates of TD fractions in young stellar clusters \cite[e.g., see Figure 11 in][]{2014prpl.conf..497E}. Interestingly, the TDs are among the strongest and largest continuum sources in our sample (Figures~\ref{fig-gallery} \& \ref{fig-g2d}). When considering only the brightest half of our continuum-detected Lupus sample, the fraction of TDs is 19\% (6/31), which is consistent with the 20\% value found for Taurus and Ophiuchus disks in \cite{2011ApJ...732...42A} (see their Figure 10). Thus on timescales of just a few Myr, a large fraction of the most massive disks show clear evidence for substantial disk evolution in the regions most relevant for planet formation.

Additionally, we found that $^{13}$CO emission is detected toward all the continuum-identified TDs, while the overall detection rate of $^{13}$CO in our continuum-detected sample is only 56\%. C$^{18}$O emission is also detected toward all our TDs with resolved cavities, while the overall detection rate of C$^{18}$O in our continuum-detected sample is only 18\%. The gas emission is clearly present {\it inside} the dust cavities of the resolved TDs (Figure \ref{fig-zoo}), consistent with previous TD observations \citep{2012ApJ...753...59M, 2013Sci...340.1199V,2013Natur.493..191C,2014A&A...562A..26B,2014ApJ...791...42Z,2015A&A...579A.106V}. The larger CO isotopologue detection rate toward the TDs may be explained by the directly heated CO wall increasing the flux of optically thick lines \citep{2013A&A...559A..46B}, or could be related to their relatively stronger continuum emission. The Lupus TDs will be fully analyzed in a separate paper (van der Marel et al., in prep.).

%===============================  DISCUSSION ============================

\section{DISCUSSION}
\label{sec-discussion}

\subsection{Correlations with Stellar Mass \label{sec-correlation}}

\capstartfalse
\begin{figure}
\begin{centering}
\includegraphics[width=8.5cm]{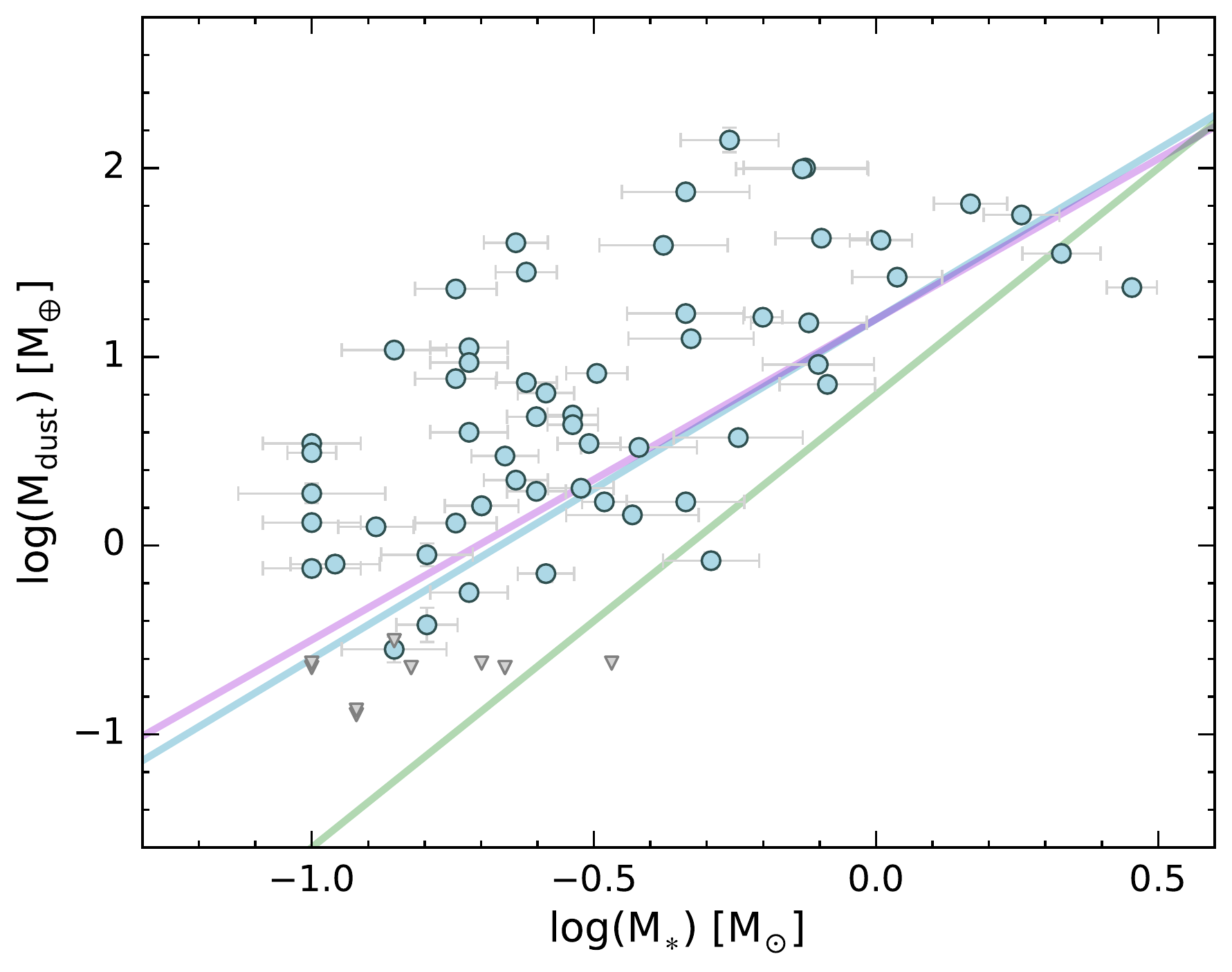}
\caption{\small Disk dust mass ($M_{\rm dust}$) as a function of stellar mass ($M_{\ast}$) for Lupus disks (Section~\ref{sec-correlation}). Blue circles show continuum detections and gray triangles show 3$\sigma$ upper limits (error bars include the 10\% absolute flux calibration error). The 20 obscured sources, for which we did not derive $M_{\ast}$ values, are not shown. The blue line gives our Bayesian linear regression fit for Lupus, while the purple and green lines show our derived relations for Taurus and Upper Sco, respectively.}
\label{fig-stellar}
\end{centering}
\end{figure}
\capstartfalse

A positive correlation between $M_{\rm dust}$ and $M_{\ast}$ has been found in sub-mm surveys of star-forming regions with increasing confidence over the last several decades (see discussion in \citealt{2013ApJ...771..129A}). The first statistically robust confirmation of this correlation was by \cite{2013ApJ...771..129A} for Class II disks in the young ($\sim$1--2~Myr) Taurus region. Barenfeld et al. (submitted) later used ALMA to derive the relation for ``primordial" disks in the older ($\sim$5--10 Myr) Upper Sco region. 

To characterize this relation in Lupus, we employed the same Bayesian linear regression technique from \cite{2007ApJ...665.1489K} used by \cite{2013ApJ...771..129A} and Barenfeld et al. (submitted). This technique characterizes linear correlations given measurement errors, upper limits, and intrinsic scatter. For the 20 obscured sources in our sample without stellar masses (Section~\ref{sec-sample}), we took a Monte Carlo (MC) approach by randomly assigning $M_{\ast}$ values based on the distribution of Lupus I--IV YSOs derived in \cite{2011MNRAS.418.1194M} (see their Figure 9). We combined the posterior distributions of 100 MC runs, finding a positive relation between ${\rm log}(M_{\rm dust})$ and ${\rm log}(M_{\ast})$ with a slope of $1.8\pm0.4$ and dispersion of $0.9\pm0.2$~dex, as shown in Figure~\ref{fig-stellar}. To confirm the significance of this relation we used the Cox proportional hazard test for censored data, implemented in the \texttt{R Project for Statistical Computing} \citep{R-Core-Team:2015aa,survival-package}, finding $<$0.0005 probability of no correlation for all MC runs. Note that we find the same results to within errors when simply removing the sources with unknown stellar masses. We also tested our results using stellar masses derived from the evolutionary models of \cite{1998A&A...337..403B} and \cite{2015A&A...577A..42B}; we found that the derived relations remain significant and consistent to within errors regardless of the model used. 

\capstartfalse
\begin{figure}
\begin{centering}
\includegraphics[width=8.5cm]{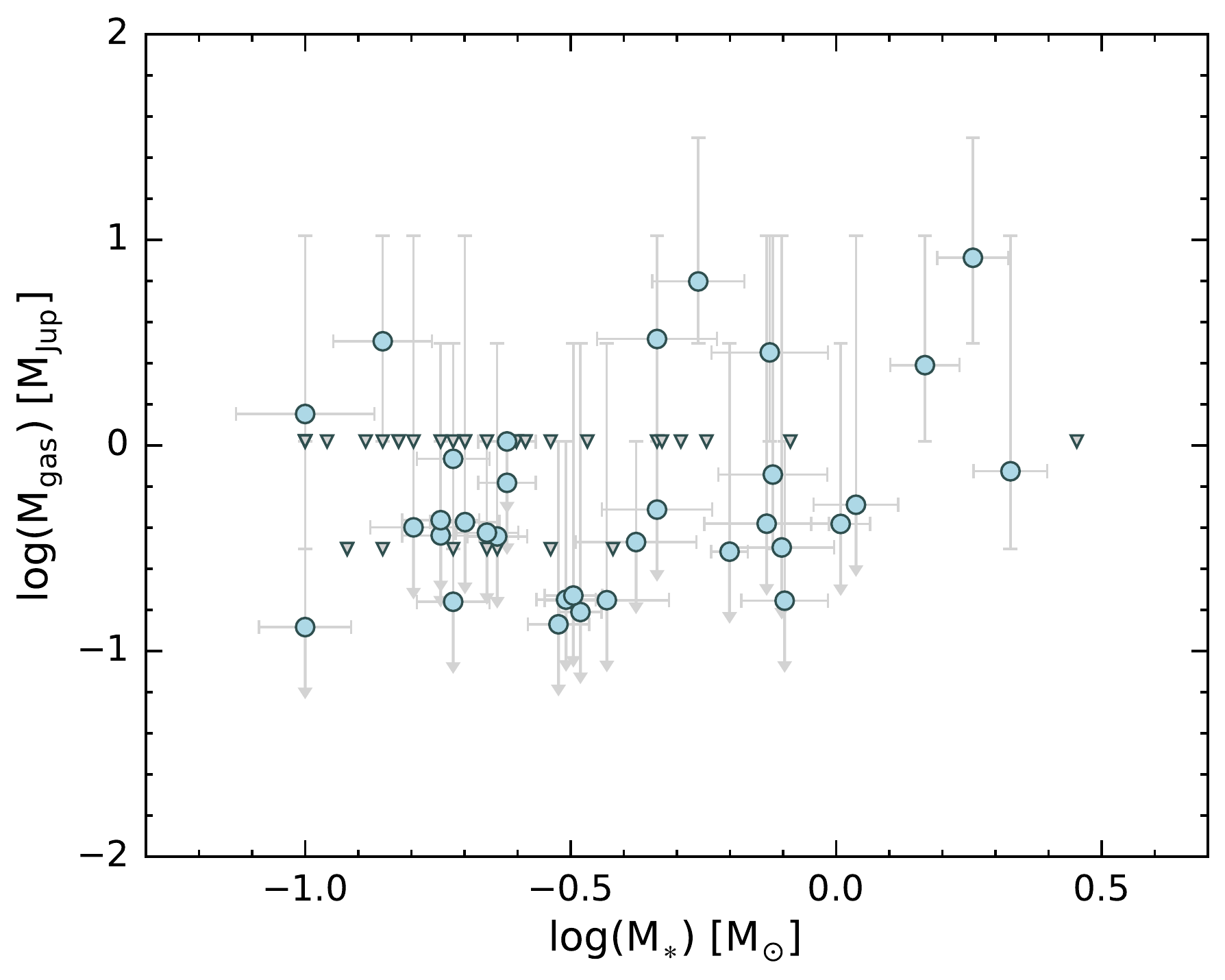}
\caption{\small Disk gas mass ($M_{\rm gas}$) as a function of stellar mass ($M_{\ast}$) for Lupus disks (Section~\ref{sec-correlation}). Blue circles show sources detected in C$^{18}$O and/or $^{13}$CO, while triangles show upper limits for sources undetected in both lines. Error bars cover the range of model gas masses, as described in Section~\ref{sec-gas}, where error bars with downward arrows represent sources with $^{13}$CO but not C$^{18}$O detections. The 20 obscured sources, for which we did not derive $M_{\ast}$ values, are not shown.}
\label{fig-stellar-gas}
\end{centering}
\end{figure}
\capstartfalse

To compare our correlation between $M_{\rm dust}$ and $M_{\ast}$ in Lupus to those found in Taurus and Upper Sco, we calculated the dust masses uniformly across each region by translating the sub-mm continuum fluxes (or $3\sigma$ upper limits) given in \cite{2013ApJ...771..129A} and Barenfeld et al. (submitted) into dust masses using Equation~\ref{eqn-mass} scaled to the distances of the clusters and the observation wavelengths of the surveys. We assumed $T_{\rm dust}=20$~K for all disks and  adopted distances of 145~pc for Upper Sco \citep{1999AJ....117..354D} and 140~pc for Taurus \citep{2008hsf1.book..405K}. We considered only sources with $M_{\ast}\gtrsim0.1~M_{\odot}$ as this was the stellar mass limit common to the three surveys. All stellar masses were derived using \cite{2000A&A...358..593S} evolutionary models. Compared to Lupus, we find a similar slope ($1.7\pm0.2$) and dispersion ($0.7\pm0.1$) for Taurus disks, but a steeper slope ($2.4\pm0.4$) and similar dispersion ($0.7\pm 0.1$) for Upper Sco disks. 

The $M_{\rm dust}$ vs. $M_{\ast}$ relations for Lupus, Taurus, and Upper Sco are shown together in Figure~\ref{fig-stellar}. The nearly identical relations for the similarly aged Taurus and Lupus regions suggests there may be a universal correlation between $M_{\rm dust}$ and $M_{\ast}$ imprinted at disk formation, while their divergence from the older Upper Sco region at lower stellar masses may indicate that disk evolution serves to steepen this initial relation with age. For example, if the most massive disks tend to have large inner holes (as suggested by this and other sub-mm surveys of star-forming regions; Section~\ref{sec-TD}) then their dust may be trapped in the outer disk for longer timescales compared to lower-mass disks, effectively steepening the relation between $M_{\rm dust}$ and $M_{\ast}$ with age. 

Confirming a clear positive correlation between $M_{\rm dust}$ and $M_{\ast}$ in Lupus supports the suggestion by \cite{2013ApJ...771..129A} that such a relation fundamentally explains the correlation between giant planet frequency and host star mass identified in the exoplanet population \citep{2006ApJ...649..436E,2007ApJ...670..833J,2013A&A...549A.109B,2010ApJ...709..396B}. This is because core growth is more efficient both in higher-mass disks \cite[e.g.,][]{2008Sci...321..814T,2012A&A...541A..97M} and around higher-mass stars \cite[e.g.,][]{2008ApJ...673..502K}. Thus the sources in the upper right of Figure~\ref{fig-stellar} are more likely to form giant planet cores before the gas disk dissipates, allowing the cores to accrete substantial gaseous envelopes and become gas giant planets. 
 
Lupus disks also exhibit a large dispersion in $M_{\rm dust}$, spanning $\sim$2 orders of magnitude for a given $M_{\ast}$. \cite{2013ApJ...771..129A} noted that the similarly large dispersion in Taurus is likely a consequence of the inherent diversity in disk temperatures, dust opacities, and evolutionary states across the region. The large dispersion in Lupus may result from different environments and/or evolutionary states across the Lupus~I--IV clouds. To test this, we re-derived the $M_{\rm dust}$ vs. $M_{\ast}$ correlation using only Lupus~III sources; we chose Lupus~III because it contains the most sources in our sample and is more distant compared to the other clouds, thus possibly in a different environment. We found consistent slope ($1.4\pm0.5$) and dispersion ($1.1\pm0.3$) values to within errors, indicating that the large dispersion in Lupus is an intrinsic property of the disk population due to the range of initial disk conditions (e.g., core angular momentum) and suggesting a range of possible planetary outcomes. 

Because we estimated the gas mass of each disk independently from the dust (Section~\ref{sec-gas}), we are able to show for the first time that $M_{\rm gas}$ and $M_{\ast}$ may also be correlated, as illustrated in Figure~\ref{fig-stellar-gas}. We used the same MC approach for assigning $M_{\ast}$ values to the 20 obscured sources without stellar masses, and again employed the Cox proportional hazard test for censored data to evaluate the significance of the correlation. We found a tentative positive correlation between $M_{\rm gas}$ and $M_{\ast}$, with a 0.01 probability of no correlation on average. However, the significant number of gas upper limits, and the large uncertainties on the gas mass estimates, means that we could not reliably determine the slope of the relation. Thus this relation should be re-visited with higher-sensitivity line observations, or in regions that have more sources with $M_{\ast} \ge 0.5~M_{\odot}$ where we detect almost all disks in gas and dust. If confirmed, this positive relation between $M_{\rm gas}$ and $M_{\ast}$ would further explain the positive correlation between giant planet frequency and host star mass seen in the exoplanet population.

\subsection{Comparison to Other Regions \label{sec-compare}}

\capstartfalse
\begin{figure}
\begin{centering}
\includegraphics[width=8.7cm]{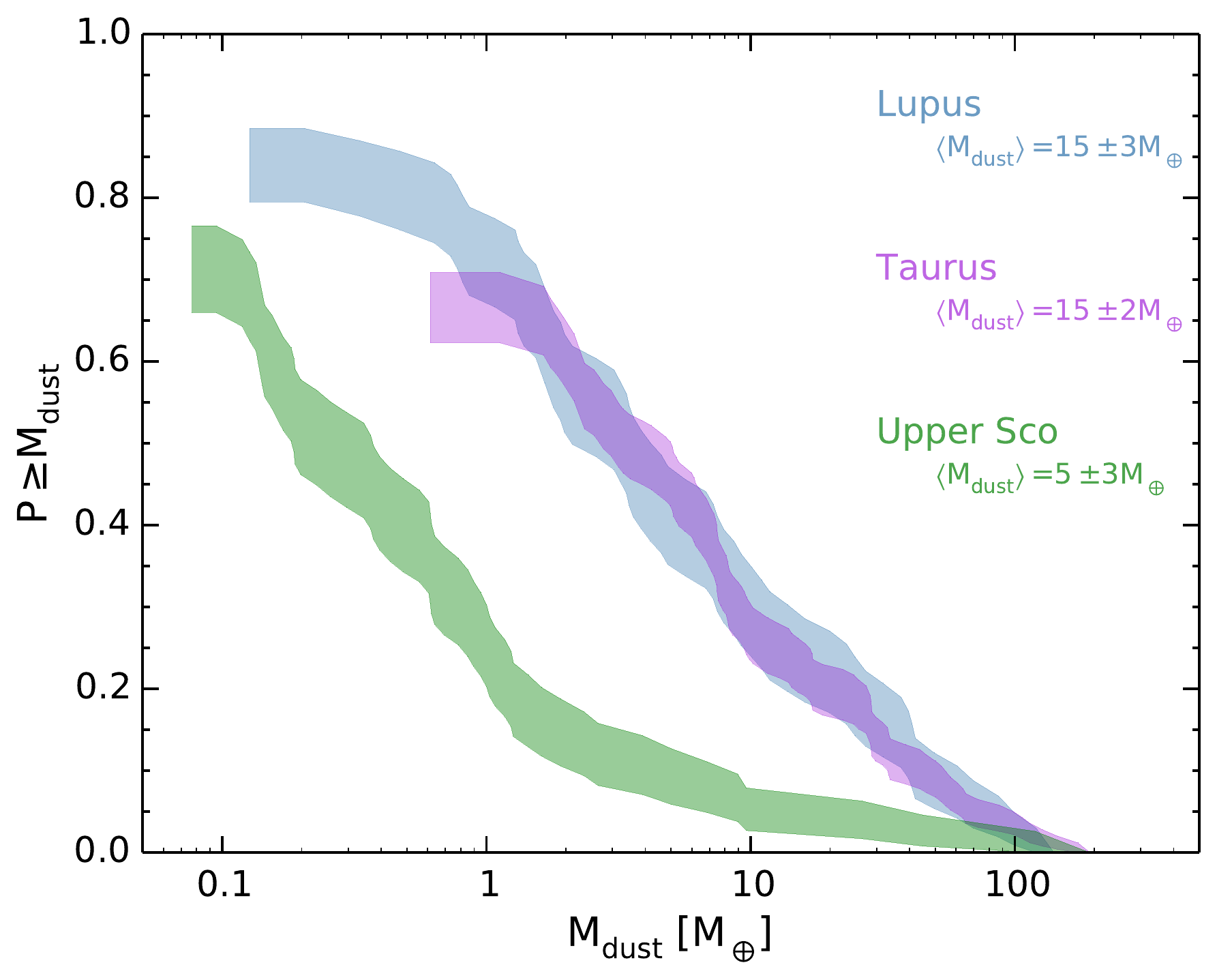}
\caption{\small Dust mass cumulative distributions for Lupus, Taurus, and Upper Sco disks around host stars with $M_{\ast}>0.1~M_{\odot}$ (Section~\ref{sec-compare}). The average dust masses for each region are given for reference. The distributions were calculated using the Kaplan-Meier estimator to include upper limits; line widths indicate 1$\sigma$ confidence intervals.}
\label{fig-compare}
\end{centering}
\end{figure}
\capstartfalse

Sub-mm surveys of star-forming regions at different ages provide the best available tool for probing dust mass evolution, as sub-mm continuum emission can be directly related to bulk dust mass (Section~\ref{sec-dust}). This work provides a near-complete census of protoplanetary disks in the young ($\sim$1--3~Myr) Lupus I--IV clouds with a dust mass sensitivity of $\sim$0.2--0.4~$M_{\oplus}$ (Section~\ref{sec-dust}), making it an ideal baseline survey of early disk conditions. In an effort to understand disk evolution, we can compare our Lupus dust mass distribution to those found in other regions located in different environments or at different stages of evolution. Only two other star-forming regions, Taurus and Upper Sco, have been surveyed in the sub-mm with similar sensitivity and completeness. Taurus has a similar age to Lupus ($\sim$1--2~Myr) and its Class II disks were surveyed down to the brown-dwarf limit with a dust mass sensitivity of $\sim$2~$M_{\oplus}$ \citep{2013ApJ...771..129A}. Upper Sco was recently surveyed in the sub-mm with ALMA with a dust mass sensitivity of $\sim$0.1~$M_{\oplus}$ (Barenfeld et al., submitted) and its older age ($\sim$5--10 Myr) makes it an important point for comparison.

We calculated dust masses uniformly across each region, as described in Section~\ref{sec-correlation}, and considered only sources with $M_{\ast}>0.1~M_{\odot}$. Figure~\ref{fig-compare} shows the dust mass cumulative distributions, calculated using the Kaplan-Meier estimator in the ASURV package \citep{1992ASPC...25..245L} to include upper limits. Lupus and Taurus have consistent mean dust masses ($15\pm3$~$M_{\oplus}$ and $15\pm2$~$M_{\oplus}$, respectively), while Upper Sco has a significantly lower mean dust mass ($5\pm3$~$M_{\oplus}$). We confirmed these results using the two-sample tests in ASURV, which estimate the probability that two samples of censored data have the same parent distribution. We found probabilities of 0.87--0.98 for Lupus and Taurus, indicating statistically similar dust mass distributions. We also found probabilities of $<5\times10^{-5}$ for Lupus and Upper Sco as well as Taurus and Upper Sco, indicating statistically different dust mass distributions.

\capstartfalse
\begin{figure}
\begin{centering}
\includegraphics[width=8.4cm]{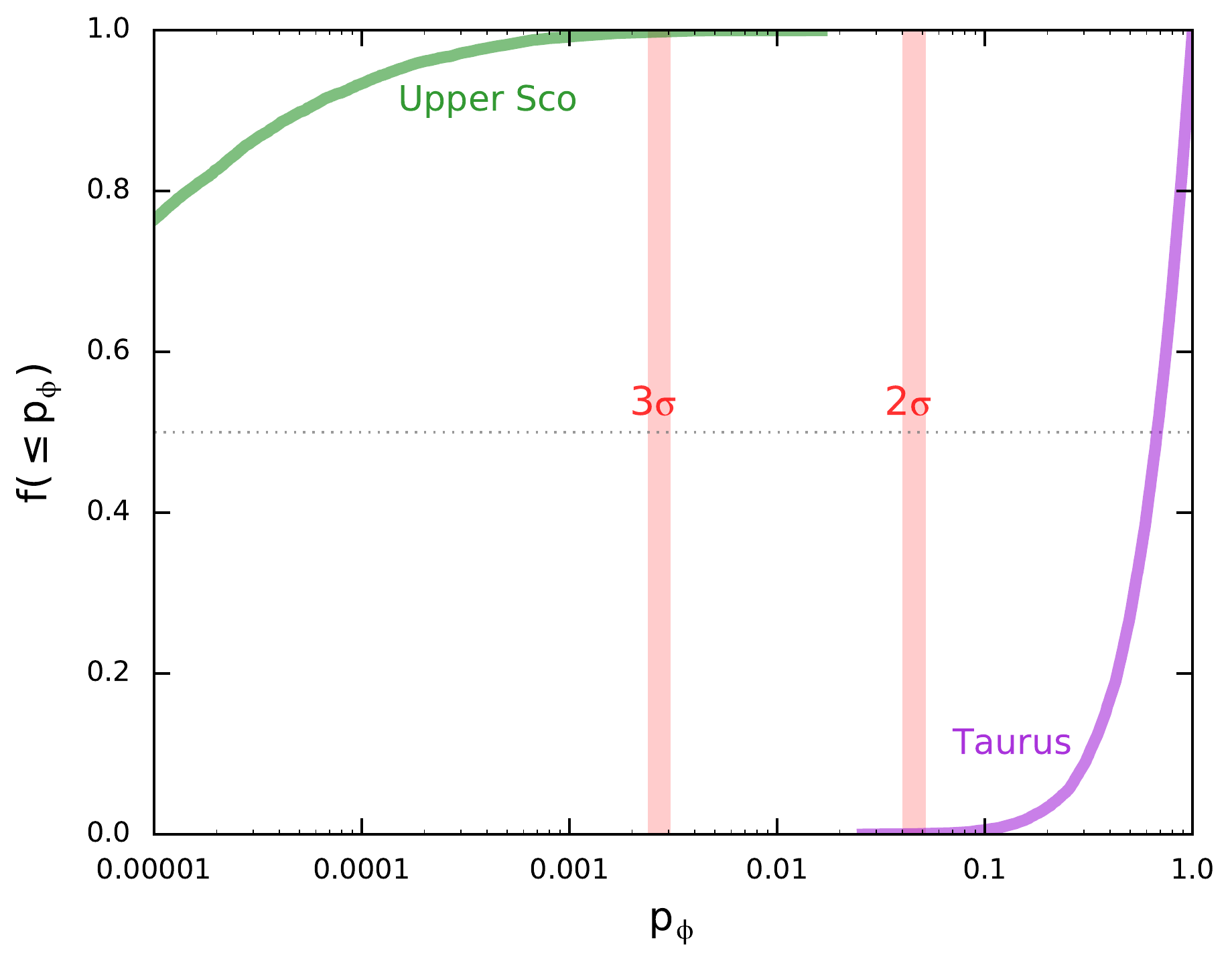}
\caption{\small Comparisons of the dust mass distributions in Taurus and Upper Sco to that of the Lupus reference sample. $p_{\phi}$ is the probability that the potentially incomplete and/or biased comparison samples are drawn from the same parent population as the complete reference sample. $f(<p_{\phi})$ is the cumulative distribution of $p_{\phi}$ constructed from two-sample tests for censored datasets in 10$^4$ MC runs. The nominal 2$\sigma$ and 3$\sigma$ probabilities that the comparison samples are different from the reference sample are shown for guidance. The median $p_{\phi}$ for Taurus is 0.67, implying a statistically indistinguishable dust mass distribution from Lupus. The median $p_{\phi}$ for Upper Sco is $1\times$10$^{-6}$, implying a statistically different dust mass distribution from Lupus.}
\label{fig-probs}
\end{centering}
\end{figure}
\capstartfalse

When comparing the dust mass distributions of two regions, it is important to confirm that they have comparable stellar distributions due to the correlation between $M_{\rm dust}$ and $M_{\ast}$ (e.g., Figure~\ref{fig-stellar}). We therefore employed the aforementioned two-sample tests in ASURV to determine the probabilities of the samples being drawn from the same parent population of stellar masses. For this analysis we removed the 20 obscured sources in our sample for which we do not have stellar masses. We found probabilities of 0.002--0.04 for Lupus and Taurus, 0.33-0.97 for Lupus and Upper Sco, and 0.0001--0.0003 for Taurus and Upper Sco. 

Thus the dust mass distribution of Lupus is readily comparable to that of Upper Sco in Figure~\ref{fig-compare}, indicating that the mean dust mass in Lupus is 3$\times$ higher than in Upper Sco. Although the dust distributions in Lupus and Taurus are remarkably similar in Figure~\ref{fig-compare}, the marginal similarities in their stellar distributions may be causing the divergence at low dust masses. Indeed, Lupus' stellar distribution is known to be dominated by late-M stars when compared to Taurus \cite[see discussion in][]{2008hsf2.book..295C}, which when combined with the correlation between $M_{\rm dust}$ and $M_{\ast}$ could at least partly explain the deviation at low dust masses. 

We also performed a more robust statistical comparison of the dust mass distributions in Taurus and Upper Sco to that of our near-complete Lupus sample, following the methodology of \cite{2013ApJ...771..129A}. Their technique uses MC simulations of two-sample test for censored datasets \citep{1985ApJ...293..192F} to take into account differences in stellar distributions, particularly mass and binarity. In short, we selected from the reference Lupus sample a subset of disks with the same distribution of host star spectral types as the comparison sample, then compared the simulated dust mass distributions using two-sample tests for censored datasets. The cumulative probability distributions of 10$^4$ MC runs are shown in Figure~\ref{fig-probs}, illustrating that the Taurus dust mass distribution is statistically indistinguishable from that of Lupus, while the Upper Sco dust mass distribution is significantly different from that of Lupus, regardless of any differences in their stellar distributions. We found consistent results when using Taurus as the reference sample instead of Lupus.

\subsection{Gas Depletion \label{sec-lowgas}}

Protoplanetary disks presumably form with an inherited ISM gas-to-dust ratio of $\sim$100 \citep{1978ApJ...224..132B}, but evolve to the opposite extreme of dusty debris disks with negligible gas in $\lesssim$10 Myr \citep{2011ARA&A..49...67W}. How quickly the dust and gas disperse, in particular relative to each other, likely dictates the types of planets that will form in a given disk. Determining the statistical properties of the dust and gas content in large samples of protoplanetary disks is therefore important for providing constraints on planet formation theories and explaining trends that are observed in the exoplanet population. 

Figure~\ref{fig-g2d} shows the dust masses, gas masses, and gas-to-dust ratios for the 62 disks in our Lupus sample that were detected in the 890~$\mu$m continuum. This represents the largest collection of disk dust and gas masses to date, providing new constraints on disk evolution. We find that, despite their moderate age of $\sim$1--3 Myr, typical disks in Lupus have gas masses well below the MMSN and gas-to-dust ratios lower than the inherited ISM value. This implies that giant planet formation is rapid, being largely complete after just a few Myr.

Additionally, such rapid gas depletion in typical protoplanetary disks may explain, at least qualitatively, the scarcity of gas giants and prevalence of intermediate-mass planets seen in the exoplanet population. In particular, exoplanet surveys have found that intermediate-mass planets (e.g., ``super-Earths" with masses between that of Earth and Neptune) are over an order of magnitude more abundant than gas giants around G/K-type stars with $P<100$~days \citep{2012ApJS..201...15H,2013ApJ...770...69P,2014PNAS..11112655M}. This finding challenges traditional planet formation theories, which predict a ``planetary desert" at intermediate masses \citep{2004ApJ...604..388I}. This is because cores of $\sim$10~$M_{\oplus}$ should have sufficient gravity to rapidly accrete gaseous envelopes, reaching masses of $\sim$1~$M_{\rm Jup}$ within $\sim$0.1~Myr if gas is still present in the disk \cite[e.g.,][]{1996Icar..124...62P}. However, if typical disks are already depleted in gas at a few Myr, such cores capable of accreting gaseous envelopes would more often end up as intermediate-mass super-Earths or Neptunes rather than gas giants. Furthermore, the fact that the exoplanet population is more of a ``tropical rainforest" at these intermediate masses \cite[i.e., exhibiting a diversity of compositions;][]{2011Natur.480..302H} may be due to the inherent diversity and rapid evolution of circumstellar disks having significant influences on the assembly of planetary systems.

Disks are stratified with gas-rich atmospheres and dust settling toward the midplane \citep{2006ApJ...638..314D}. This may be the root cause of the preferential loss of gas relative to dust via photoevaporation \citep{2014prpl.conf..475A}, layered accretion \citep{1996ApJ...457..355G}, and/or disk winds \citep{2015ApJ...801...84G,2016ApJ...818..152B}. Rapid gas depletion would also be consistent with the findings of \cite{2010A&A...510A..72F}, who used spectroscopically measured accretion rates (i.e., a completely different methodology from this work) to show that inner gas disk lifetimes are shorter than inner dust dissipation timescales; here we extend this finding to disk-averaged values. 

The main caveat with our derived gas masses is that they depend inversely on the assumed ${\rm [CO]/[H_{2}]}$ molecular abundance and ${\rm [CO]/[^{13}CO]}$ and ${\rm [CO]/[C^{18}O]}$ isotopologue ratios. WB14 assumed an ISM-like $\rm {[CO]/[H_2]}=10^{-4}$ abundance and isotopologue ratios of ${\rm [CO]/[^{13}CO]} = 70$ and ${\rm [CO]/[C^{18}O]} = 550$ or 1650. These values are consistent with those measured in molecular clouds \citep{1982ApJ...262..590F,1994ApJ...428L..69L,2013MNRAS.431.1296R,2014A&A...564A..68S} as well as with a direct measurement in a disk \citep{2014ApJ...794..160F}. However, the strong HD \citep{2013Natur.493..644B} but weak C$^{18}$O emission toward the TW Hydra disk has been interpreted as a much lower CO abundance in this system \citep{2013ApJ...776L..38F}. \cite{2016arXiv160101449K} suggested that repeated cycling through the midplane may ``dry out" the CO from the warm molecular layer and significantly reduce the $\rm {[CO]/[H_2]}$ abundance. Such an effect would increase our inferred gas masses and gas-to-dust ratios. Our data cannot distinguish between these possibilities, but regardless of the cause, the weak CO isotopologue emission indicates rapid disk evolution, either directly in the gas-to-dust ratio or chemically via permanent loss of volatiles to solids.

Finally, it is important to note that the low gas-to-dust ratios are not due to over-estimated dust masses. For realistic conditions of grain compositions and sizes, \cite{1994A&A...291..943O} show that the dust opacity, $\kappa$, used in Equation~\ref{eqn-mass} does not change sufficiently to account for the factor of $\sim$10 discrepancy between our inferred gas-to-dust ratios and that of the ISM. If anything, the growth of planetesimals, and lock-up of solids into meter- and larger-sized bodies, would decrease the continuum emission and thereby increase the apparent gas-to-dust ratio.

%===============================  SUMMARY ============================

\section{Summary}
 \label{sec-summary}
 
We presented the first high-resolution sub-mm survey of both dust and gas for a large sample of protoplanetary disks in an effort to better understand how circumstellar disks may evolve into the observed exoplanet population. 

\begin{itemize}

\item We used ALMA to survey protoplanetary disks in the young (1--3~Myr) and nearby (150--200~pc) Lupus~I--IV clouds. The region's proximity and youth make it ideal for a baseline study of early disk properties.

\item We obtained ALMA Cycle 2 data for 89 disks in the 890~$\mu$m continuum and $^{13}$CO and C$^{18}$O 3--2 lines. We detected 62 disks in the continuum, 36 in $^{13}$CO, and 11 in C$^{18}$O. All sources detected in C$^{18}$O were detected in $^{13}$CO, and all sources detected in $^{13}$CO were detected in the continuum. 

\item The continuum emission constrained $M_{\rm dust}$ down to a few Martian masses and the CO isotopologue emission constrained $M_{\rm gas}$ down to $\lesssim$1~$M_{\rm Jup}$ (assuming ISM-like [CO]/[H$_{2}$] abundance). The dust masses spanned $\sim$3 orders of magnitude and the gas masses were typically $\lesssim1~M_{\rm Jup}$.

\item Our stacking analysis showed that the average dust mass of an undetected Lupus disk was $\lesssim$6 Lunar masses (0.03~$M_{\oplus}$), indicating that protoplanetary disks evolve rapidly to debris disk levels once disk clearing begins.

\item We derived a positive correlation between $M_{\rm dust}$ and $M_{\ast}$ for Lupus disks, with a slope and dispersion nearly identical to those of the similarly aged Taurus region. We also presented the first evidence for a positive correlation between $M_{\rm gas}$ and $M_{\ast}$. Both relations would provide an origin for the dependence of giant planet frequency on stellar mass that is seen in the exoplanet population.

\item By comparing our continuum results to sub-mm surveys of other star-forming regions, we found that the mean dust mass in Lupus is 3$\times$ higher than that of the older Upper Sco region. We also found that Lupus and the similarly aged Taurus region have consistent mean dust masses and statistically indistinguishable dust mass distributions.

\item Typical disks in Lupus have gas masses well below the MMSN and gas-to-dust ratios lower than the ISM. The inferred rapid gas depletion indicates that giant planet formation is largely complete by a few Myr, and may also explain the unexpected prevalence and diversity of intermediate-mass planets seen in the exoplanet population. Although the gas masses may be underestimated due to our assumption of an ISM-like $\rm {[CO]/[H_2]}$ abundance, the weak CO isotopologue emission indicates rapid disk evolution, either directly in the gas-to-dust ratio or chemically via permanent loss of volatiles to solids.

\end{itemize}

\begin{acknowledgements}
MCA and JPW were supported by NSF and NASA grants AST-1208911 and NNX15AC92G, respectively. MCA acknowledges student observing support from NRAO. NM is supported in part by the Beatrice W. Parrent Fellowship in Astronomy at the University of Hawaii. Leiden is supported by the European Union A-ERC grant 291141 CHEMPLAN, by the Netherlands Research School for Astronomy (NOVA), and by grant 614.001.352 from the Netherlands Organization for Scientific Research (NWO). CFM gratefully acknowledges an ESA Research Fellowship. We thank J.M. Alcal\'{a} for providing VLT/X-Shooter spectra for deriving stellar parameters, Ilse Cleeves for providing ALMA continuum and integrated line fluxes for Sz~82 (IM Lup), and Scott Barenfeld for providing ALMA continuum fluxes for Upper Sco sources in advance of publication. We also thank Sean Andrews for his extremely useful comments. This paper makes use of the following ALMA data: ADS/JAO.ALMA\#2013.1.00220.S. ALMA is a partnership of ESO (representing its member states), NSF (USA) and NINS (Japan), together with NRC (Canada), NSC and ASIAA (Taiwan), and KASI (Republic of Korea), in cooperation with the Republic of Chile. The Joint ALMA Observatory is operated by ESO, AUI/NRAO and NAOJ. The National Radio Astronomy Observatory is a facility of the National Science Foundation operated under cooperative agreement by Associated Universities, Inc.

\end{acknowledgements}

%===============================  REFERENCES ============================

\bibliography{../bib.bib}

%===============================  APPENDIX ============================

\clearpage

\appendix

\section{A. Rejected Targets}
\label{app-rejected}

Eight of the sources in our original target list were observed during our ALMA Cycle 2 program, but later found to be background giants based on the lack of Li {\sc I} absorption at 6707.8 \AA{} in the VLT/X-Shooter spectra (Alcal\'{a} et al., in prep) as well as discrepant surface gravities and radial velocities with respect to Lupus YSOs (Frasca et al., in prep). All eight sources were undetected in the continuum and line. Table~\ref{tab-rejected} gives the names of these rejected sources (Sz or 2MASS), the phase center coordinates of their ALMA observations, their 890~$\mu$m continuum fluxes derived from {\it uvmodelfit} using a point-source model (Section~\ref{sec-cont}), and their $^{13}$CO and C$^{18}$O upper limits (Section~\ref{sec-line}).  
  
\setcounter{table}{0}
\renewcommand{\thetable}{A\arabic{table}}

\begin{table}[!htb]
\caption{Rejected Targets} 
\label{tab-rejected}
\centering         
\begin{tabular}{lccrcc}
\hline\hline
Source &  RA$_{\rm J2000}$ & Dec$_{\rm J2000}$ & $F_{\rm cont}$ & $F_{\rm 13CO}$        & $F_{\rm C18O}$ \\ 
             &                                &                                 & (mJy)                & (mJy~km~s$^{-1}$)  & (mJy~km~s$^{-1}$) \\
 \hline   
Sz78                            &  15:53:41.18 & -39:00:37.10   &  0.00$\pm$0.28   & $<180$   &  $<207$  \\
Sz79                            &  15:53:42.68 & -38:08:10.40   & -0.55$\pm$0.27    & $<168$   &  $<198$  \\
J15594528-4154572   &  15:59:45.28 & -41:54:57.20   &  0.12$\pm$0.19   & $<105$   &  $<120$  \\
J16000742-4149484   &  16:00:07.43 & -41:49:48.90   & -0.10$\pm$0.18   & $<102$   &  $<120$  \\
J16070863-3947219   &  16:07:08.64 & -39:47:22.70   &  0.38$\pm$0.30     & $<189$   &  $<219$   \\
J16080618-3912225   &  16:08:06.17 & -39:12:22.50   &  0.34$\pm$0.23     & $<102$   &  $<120$  \\
J16114865-3817580   &  16:11:48.67 & -38:17:58.30   &   0.12$\pm$0.30    &  $<189$   &  $<234$  \\
J16122269-3713276   &  16:12:22.73 & -37:13:27.60   &  -0.04$\pm$0.59   &  $<171$   &  $<204$   \\
\hline    
\end{tabular}
\end{table}

\clearpage
\section{B. Continuum and Line Maps}
\label{app-zoo}

\setcounter{figure}{0}
\renewcommand{\thefigure}{B\arabic{figure}}

\begin{figure*}[!ht]
\begin{center}
\includegraphics[width=15cm]{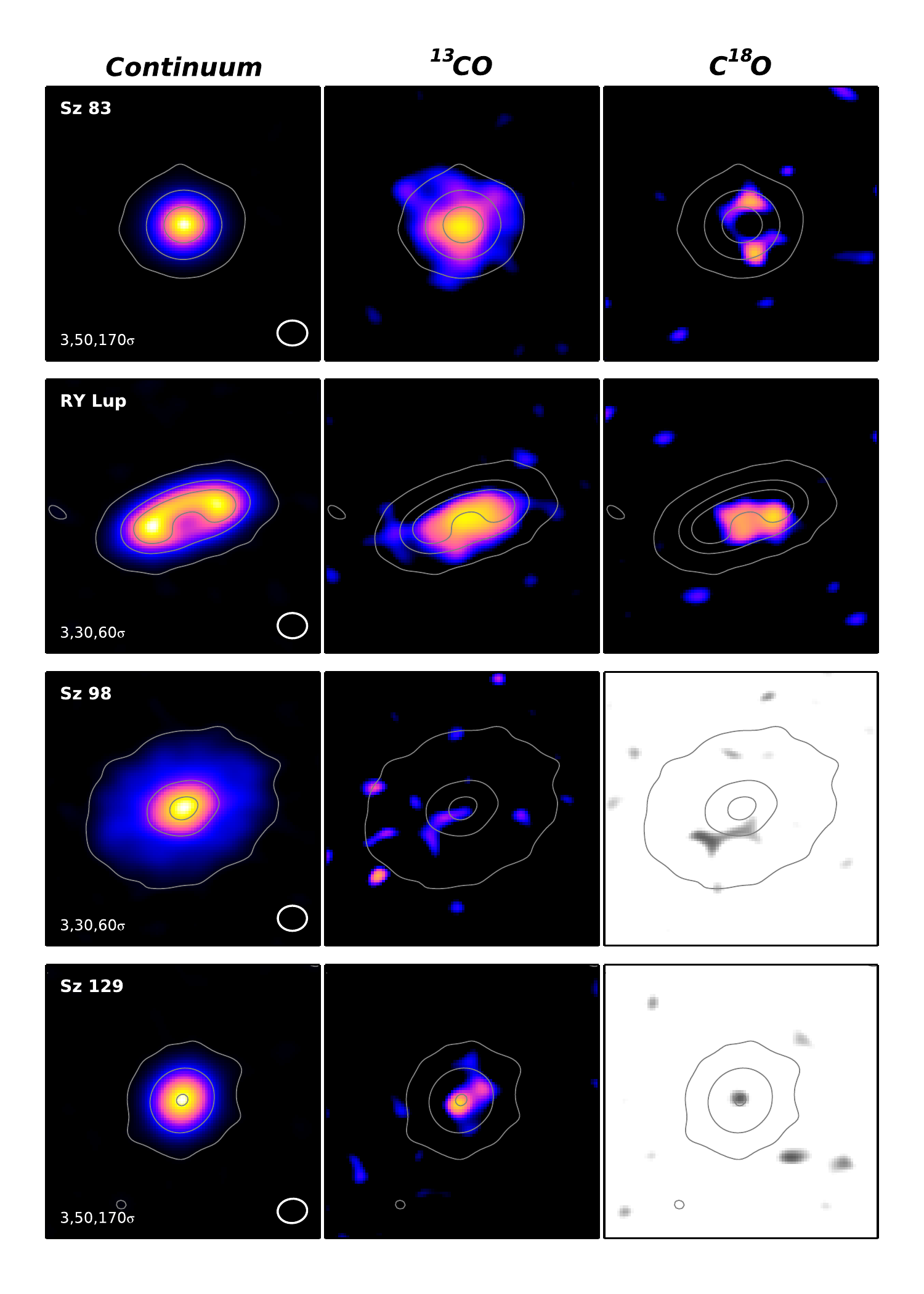}
\captcont{ALMA 890 $\mu$m continuum images (left), $^{13}$CO zero-moment maps (middle), and C$^{18}$O zero-moment maps (right) for all Lupus YSOs observed by our ALMA Cycle 2 program. Images are 3\as{}$\times$3\as{} in size. Gray images indicate non-detections. Synthesized beams are shown in the lower right corner of the continuum images.}
\label{fig-zoo}
\end{center}
\end{figure*}

%\iffalse

\begin{figure*}[!ht]
\begin{center}
\includegraphics[width=17cm]{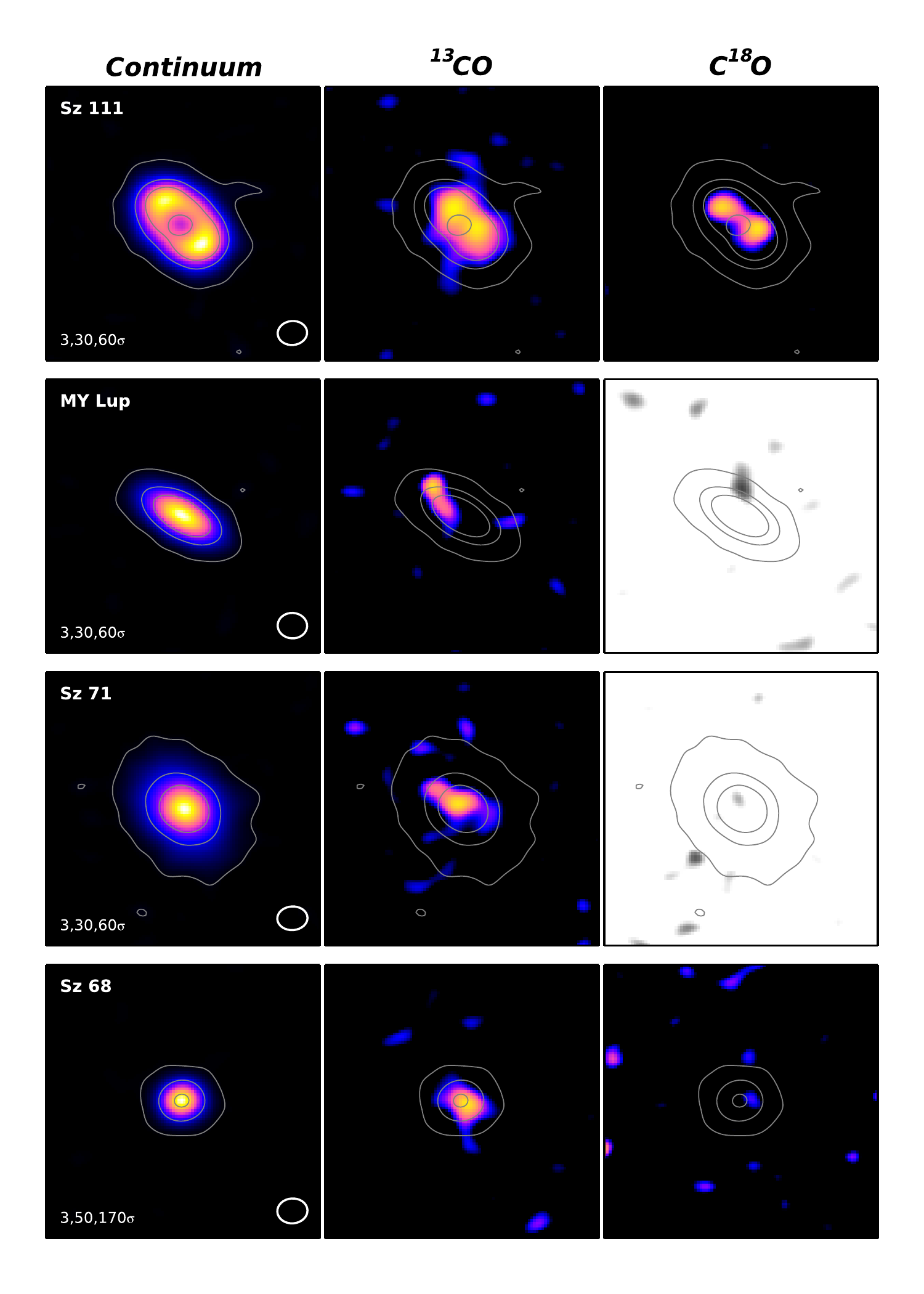}
\captcont{Cont.}
\end{center}
\end{figure*}

\begin{figure*}[!ht]
\begin{center}
\includegraphics[width=17cm]{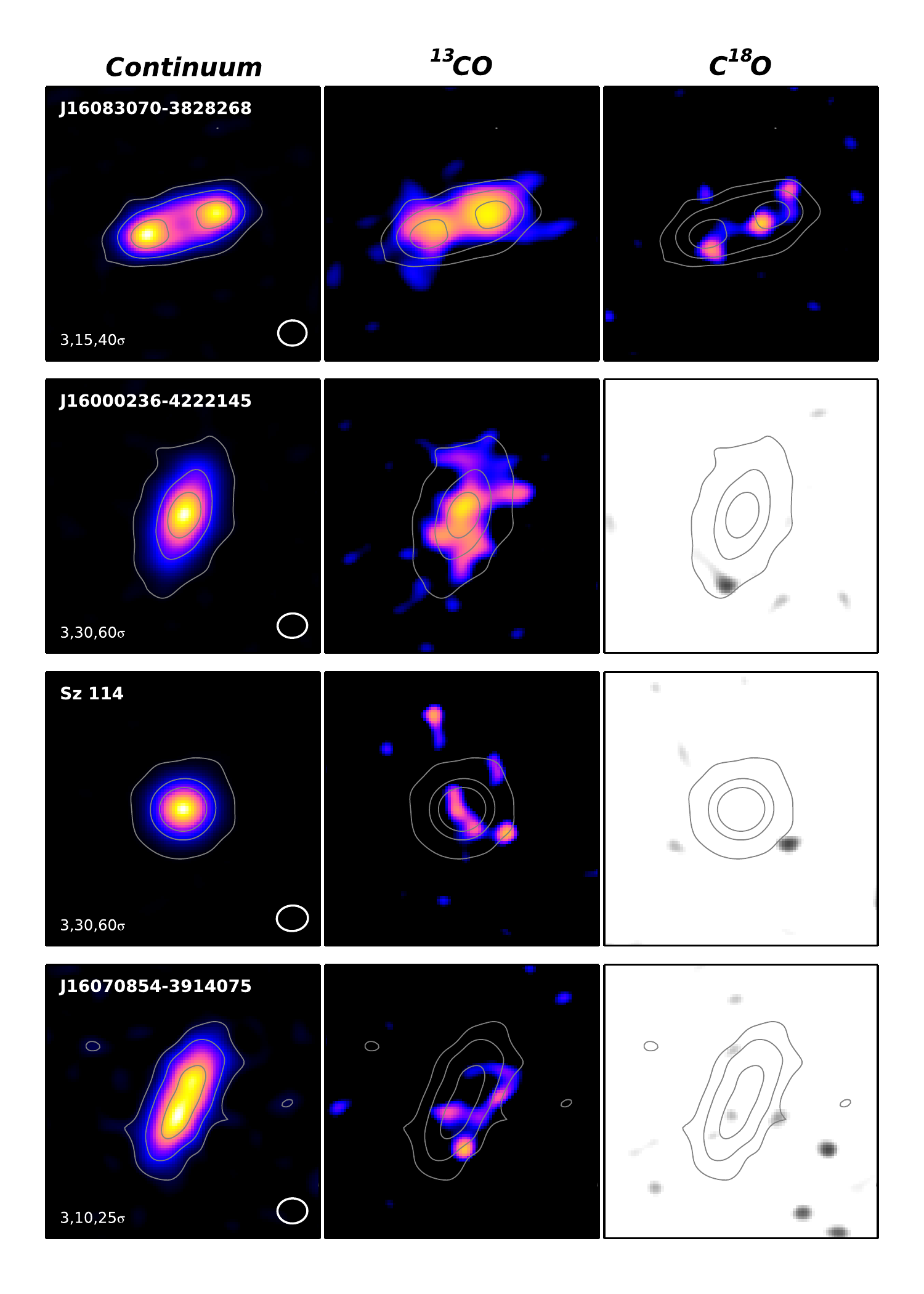}
\captcont{Cont.}
\end{center}
\end{figure*}

\begin{figure*}[!ht]
\begin{center}
\includegraphics[width=17cm]{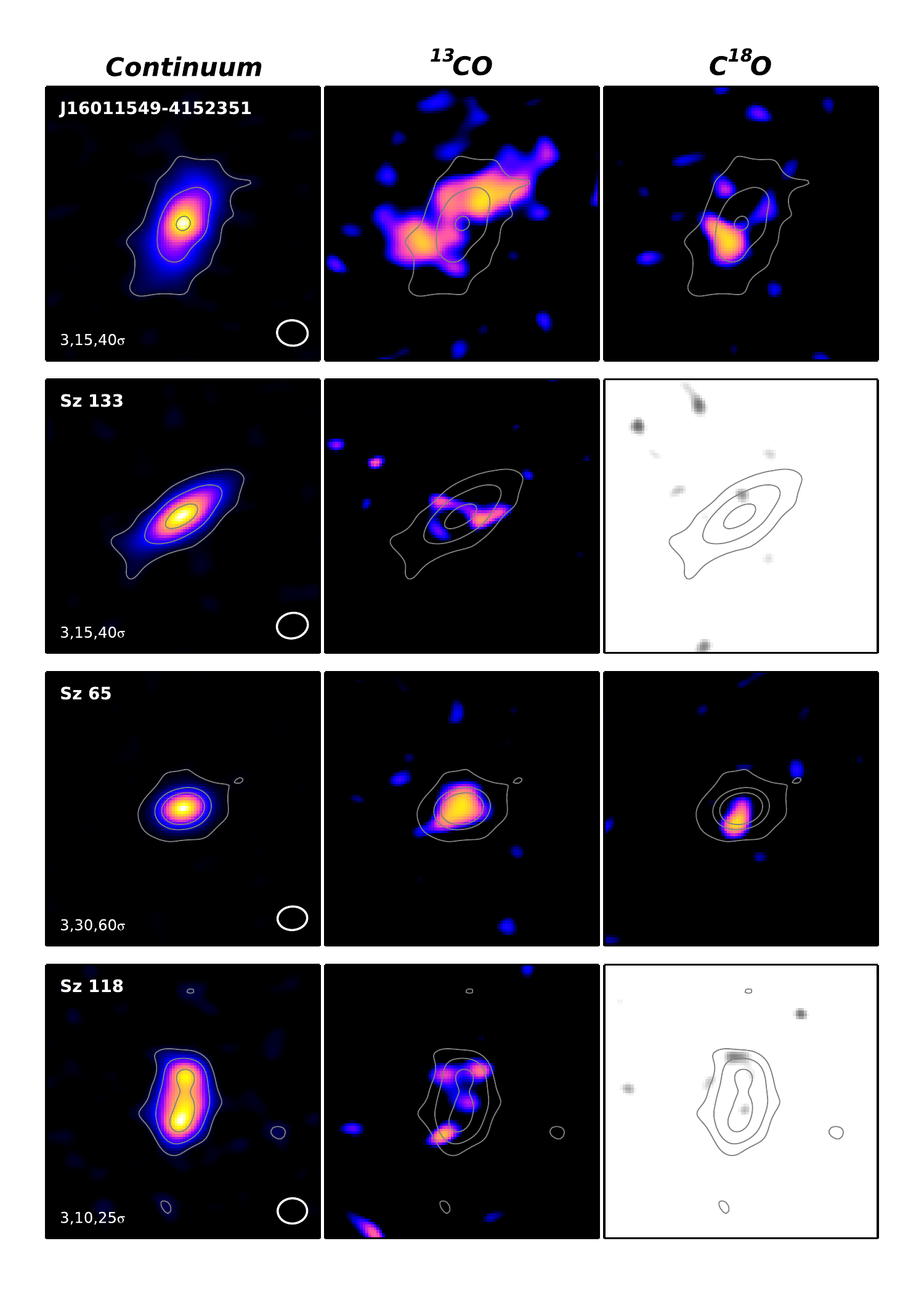}
\captcont{Cont.}
\end{center}
\end{figure*}

\begin{figure*}[!ht]
\begin{center}
\includegraphics[width=17cm]{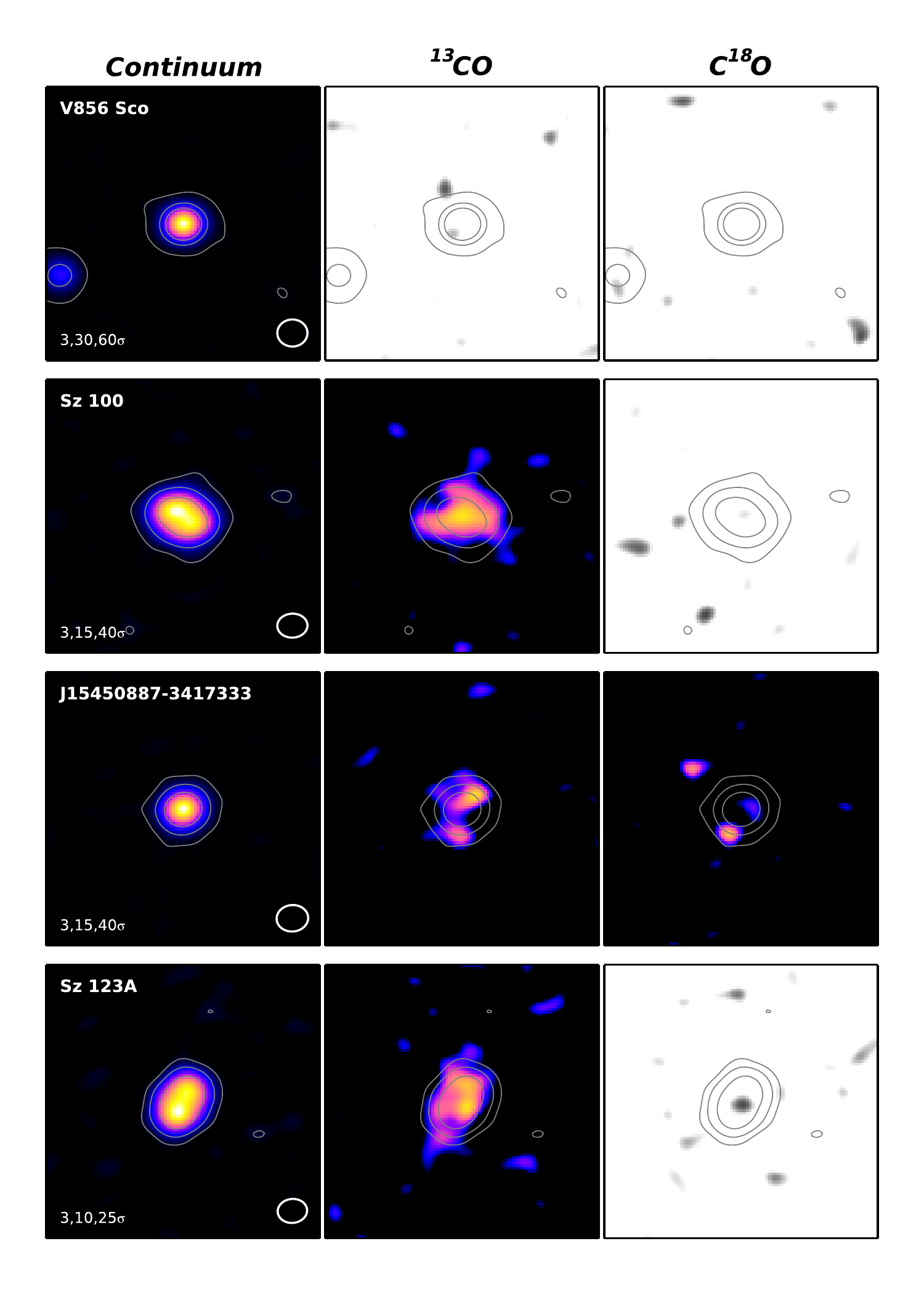}
\captcont{Cont.}
\end{center}
\end{figure*}

\begin{figure*}[!ht]
\begin{center}
\includegraphics[width=17cm]{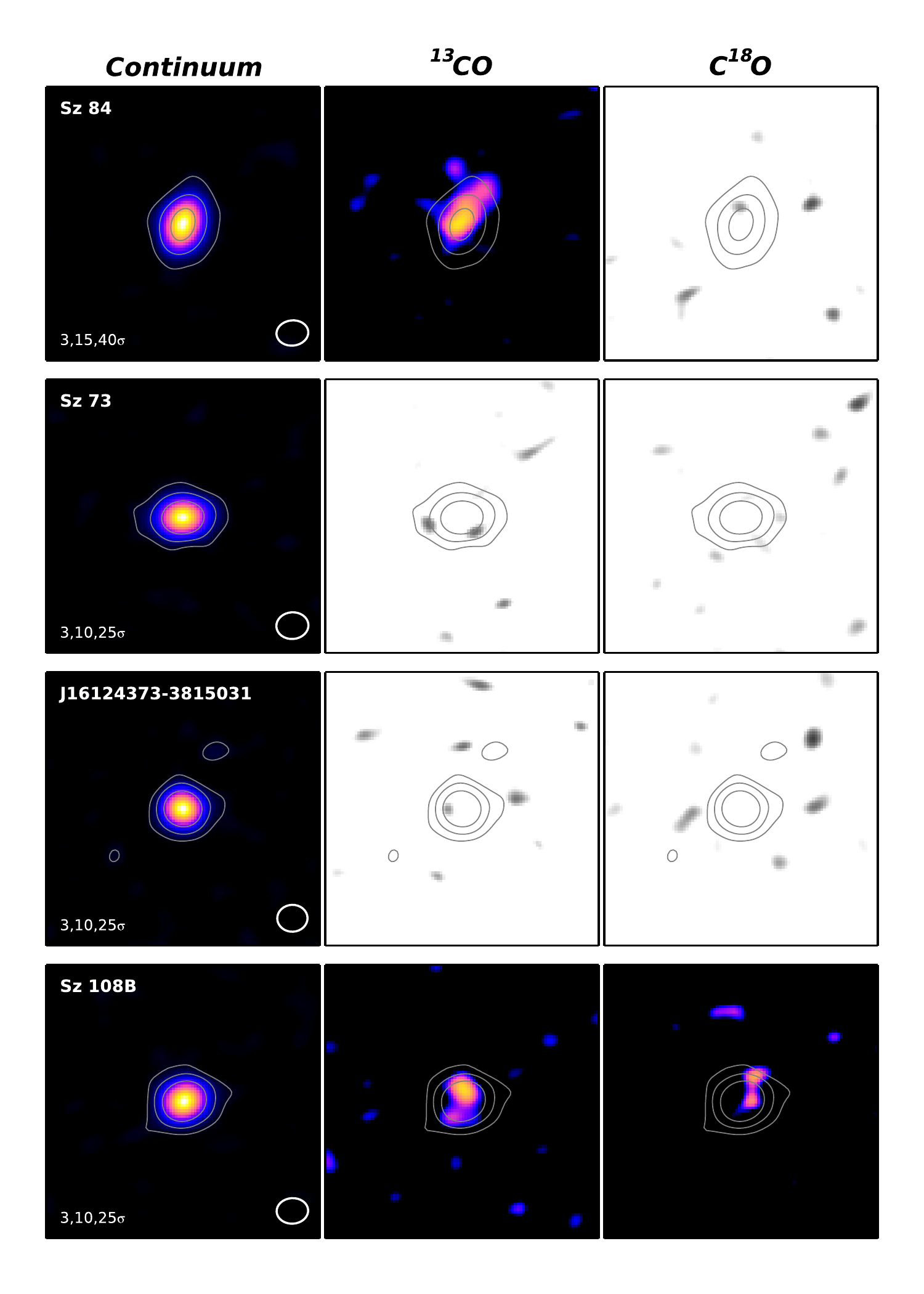}
\captcont{Cont.}
\end{center}
\end{figure*}

\begin{figure*}[!ht]
\begin{center}
\includegraphics[width=17cm]{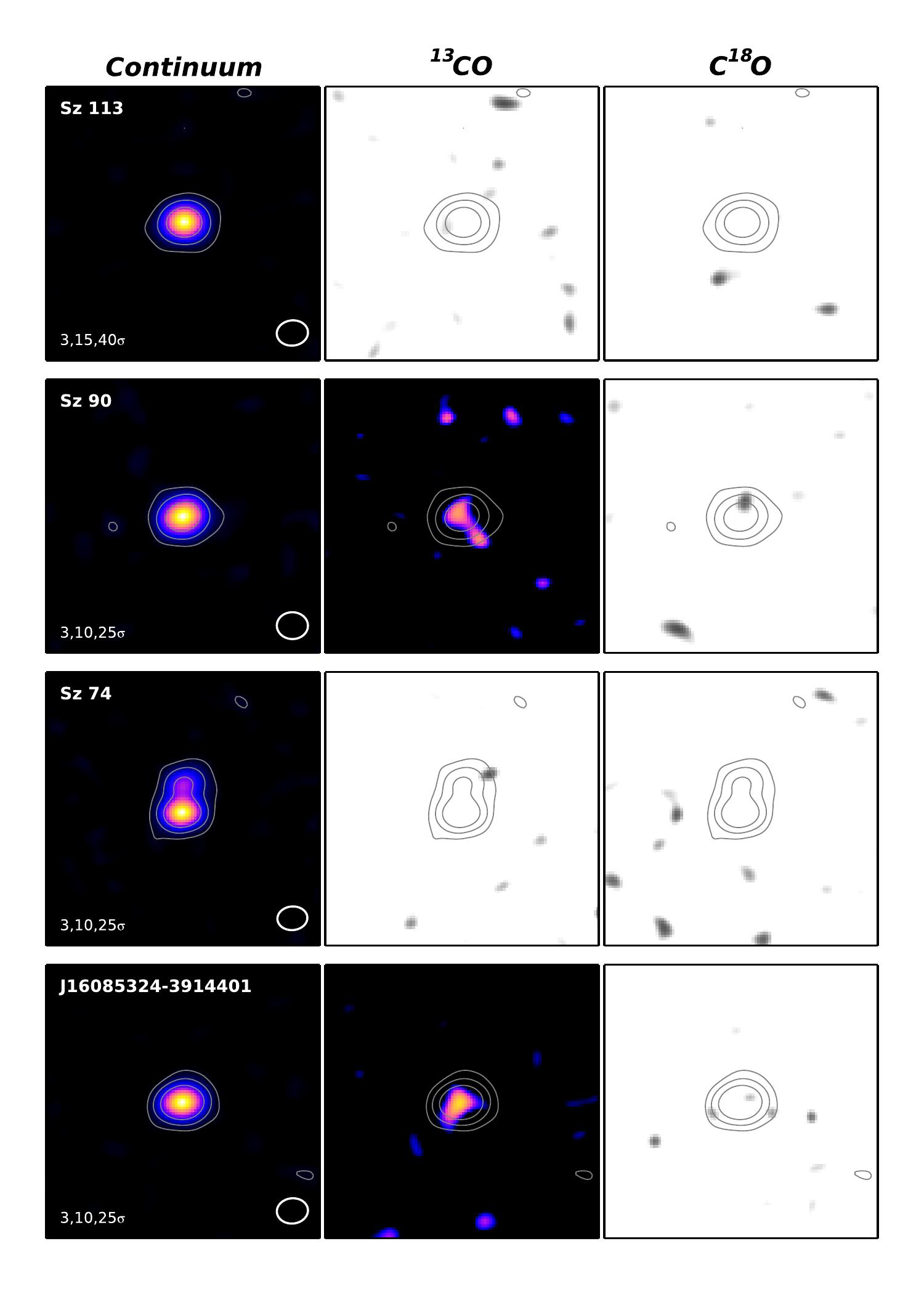}
\captcont{Cont.}
\end{center}
\end{figure*}

\begin{figure*}[!ht]
\begin{center}
\includegraphics[width=17cm]{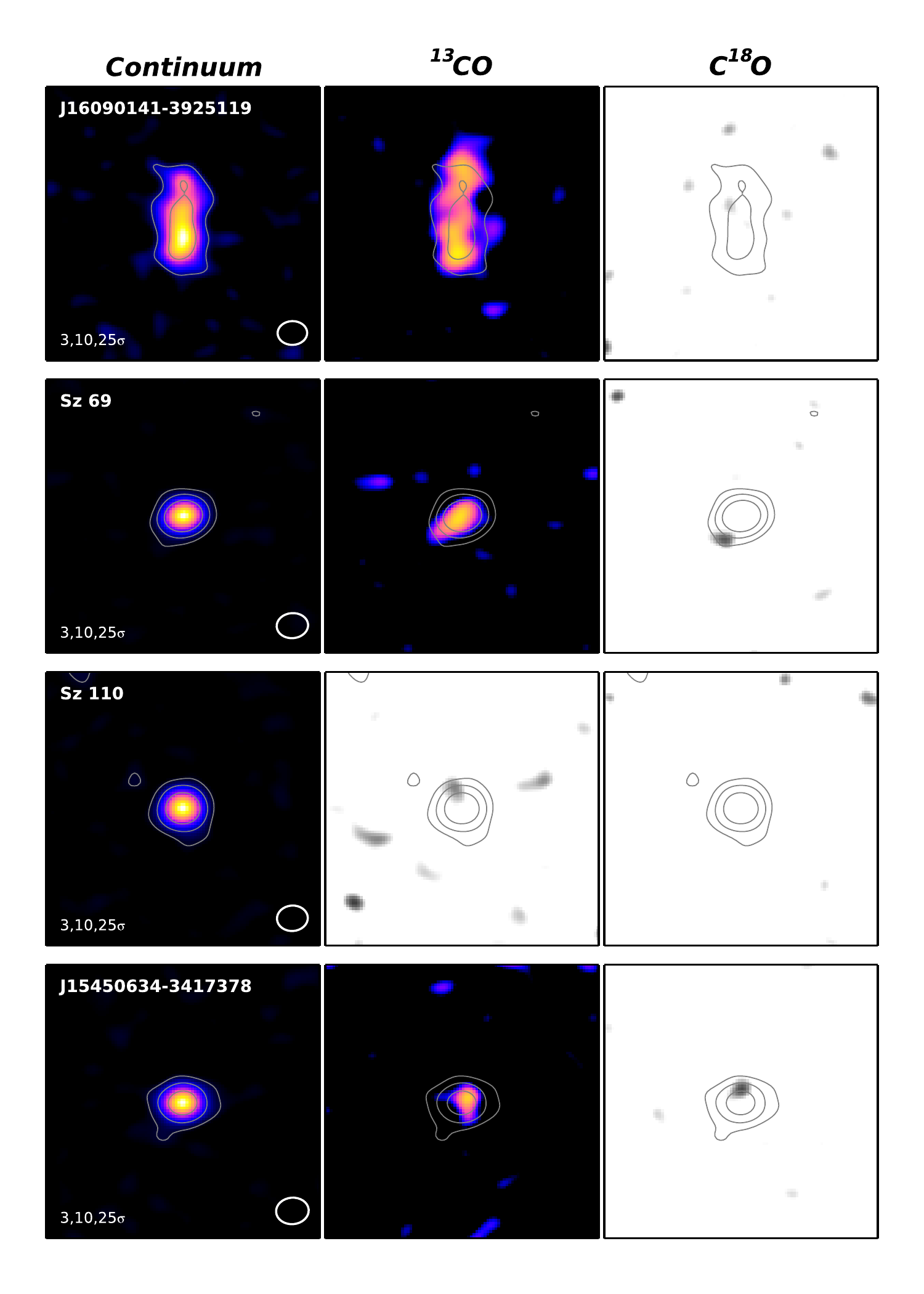}
\captcont{Cont.}
\end{center}
\end{figure*}

\begin{figure*}[!ht]
\begin{center}
\includegraphics[width=17cm]{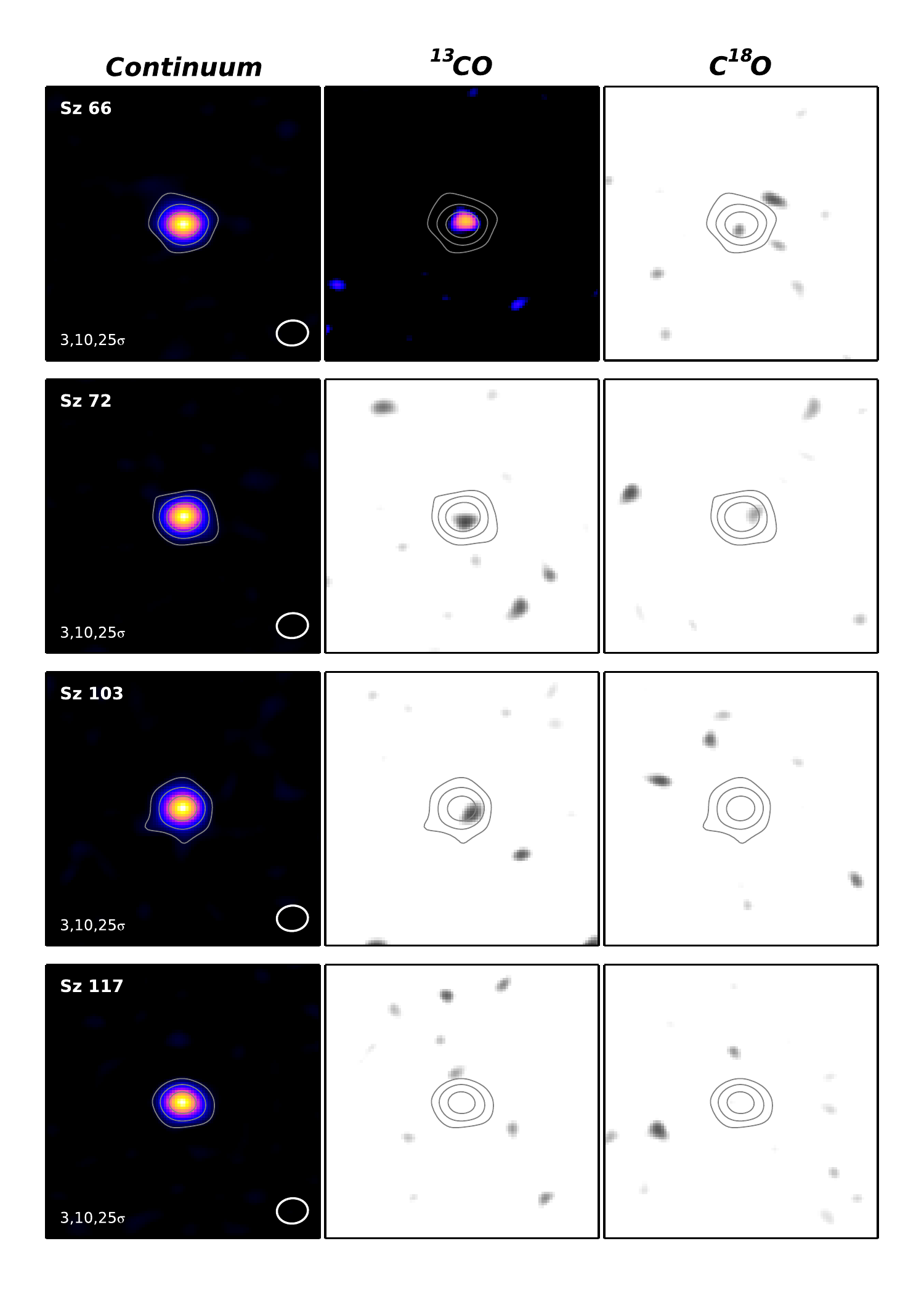}
\captcont{Cont.}
\end{center}
\end{figure*}

\begin{figure*}[!ht]
\begin{center}
\includegraphics[width=17cm]{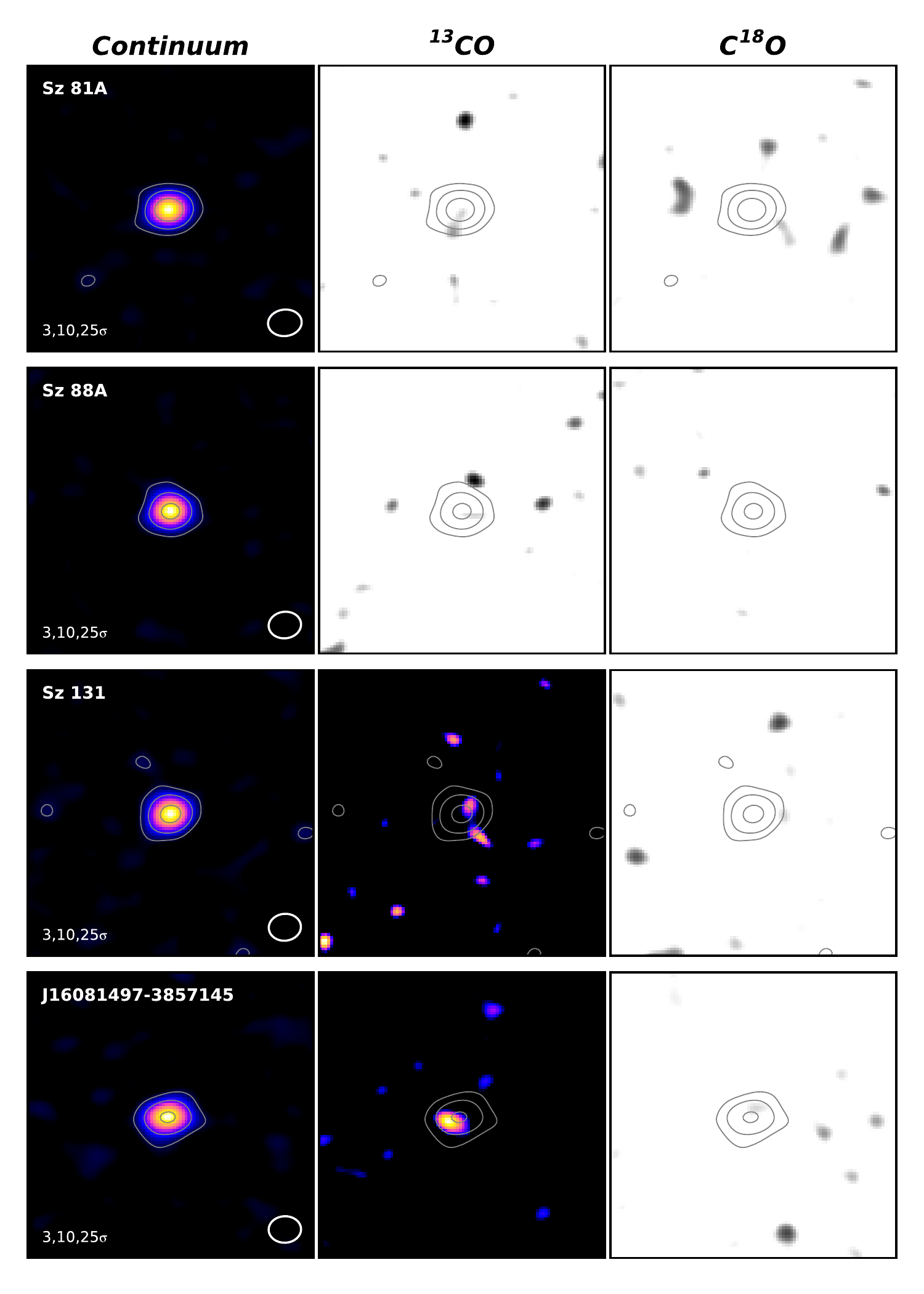}
\captcont{Cont.}
\end{center}
\end{figure*}

\begin{figure*}[!ht]
\begin{center}
\includegraphics[width=17cm]{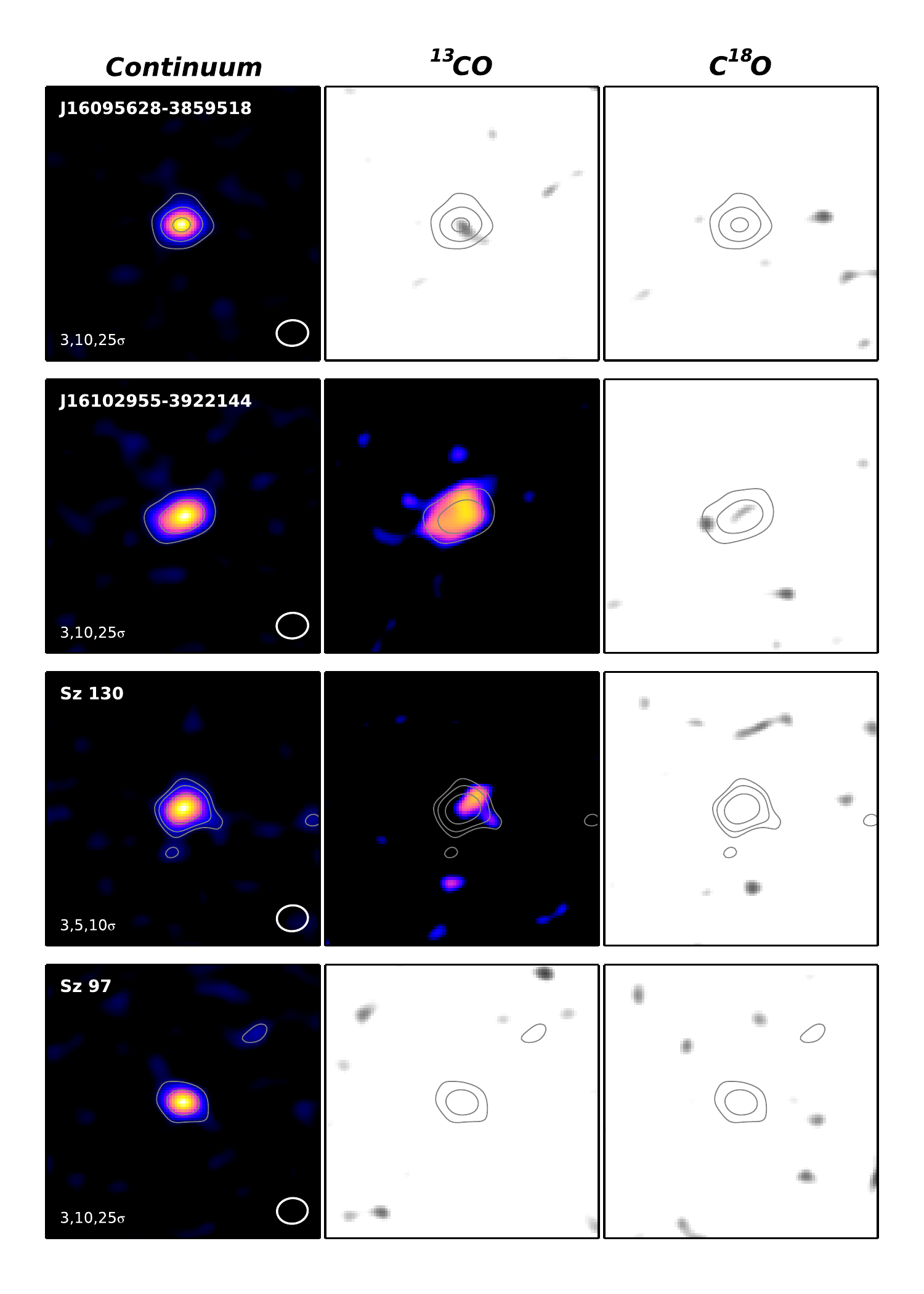}
\captcont{Cont.}
\end{center}
\end{figure*}

\begin{figure*}[!ht]
\begin{center}
\includegraphics[width=17cm]{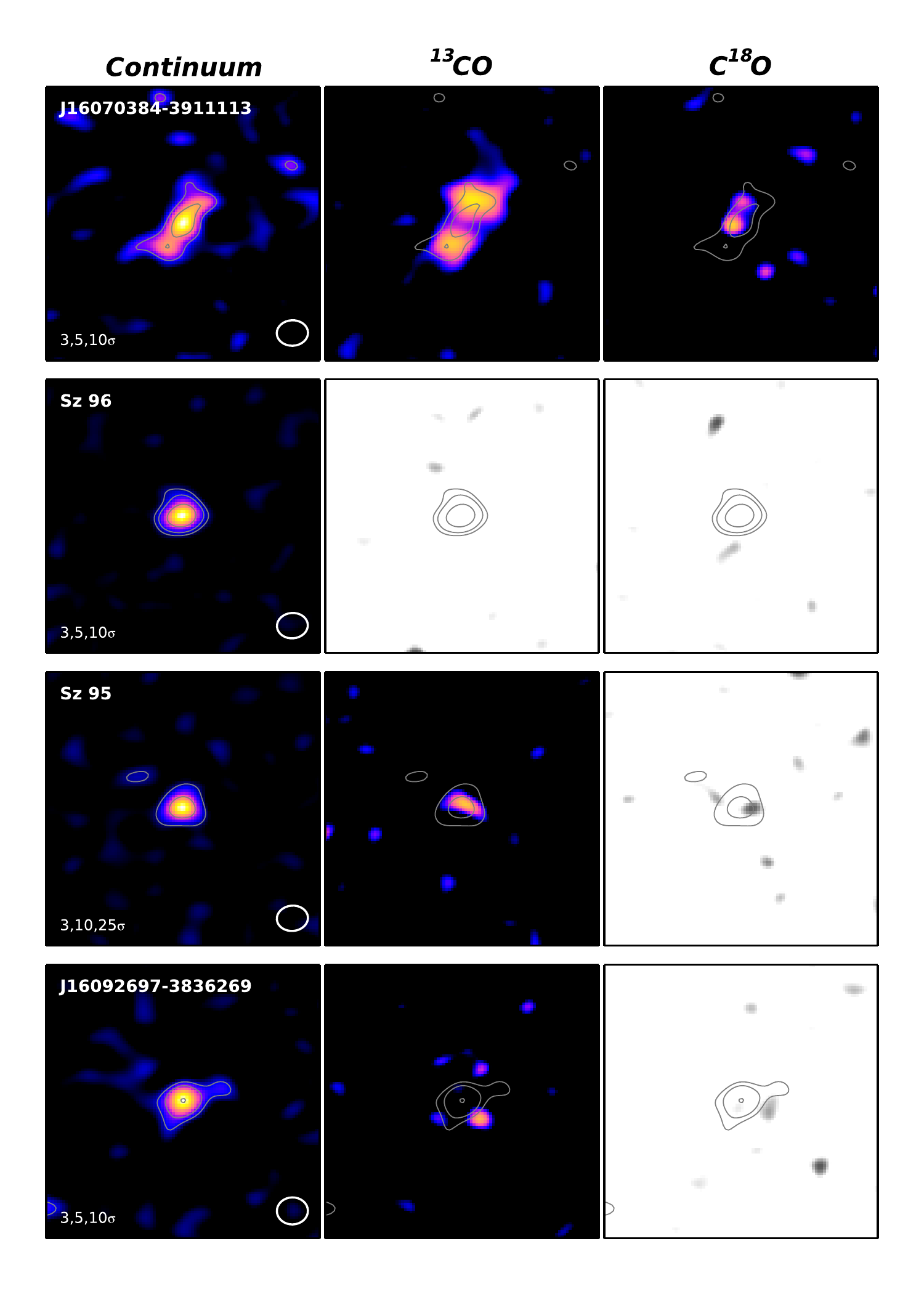}
\captcont{Cont.}
\end{center}
\end{figure*}

\begin{figure*}[!ht]
\begin{center}
\includegraphics[width=17cm]{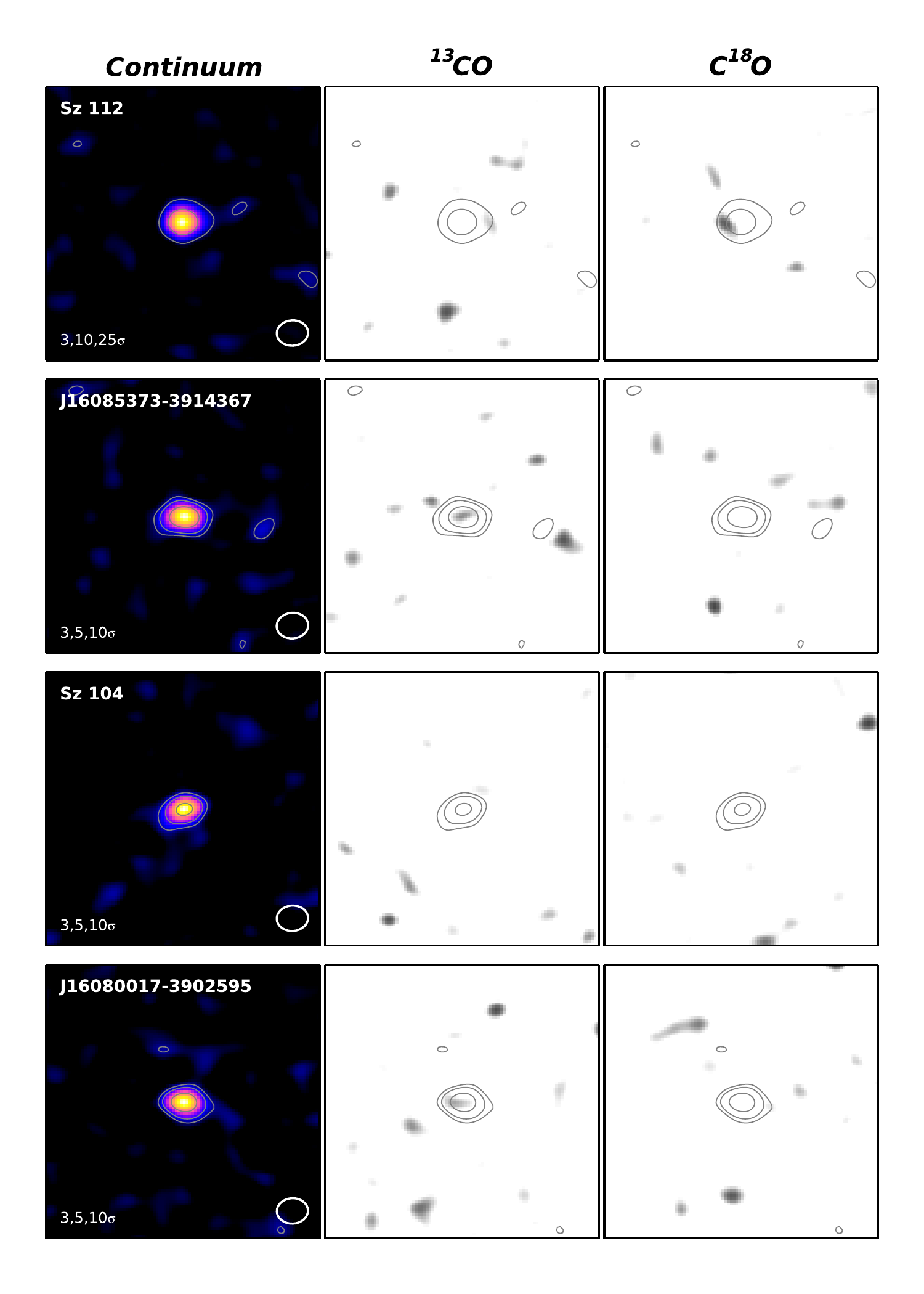}
\captcont{Cont.}
\end{center}
\end{figure*}

\begin{figure*}[!ht]
\begin{center}
\includegraphics[width=17cm]{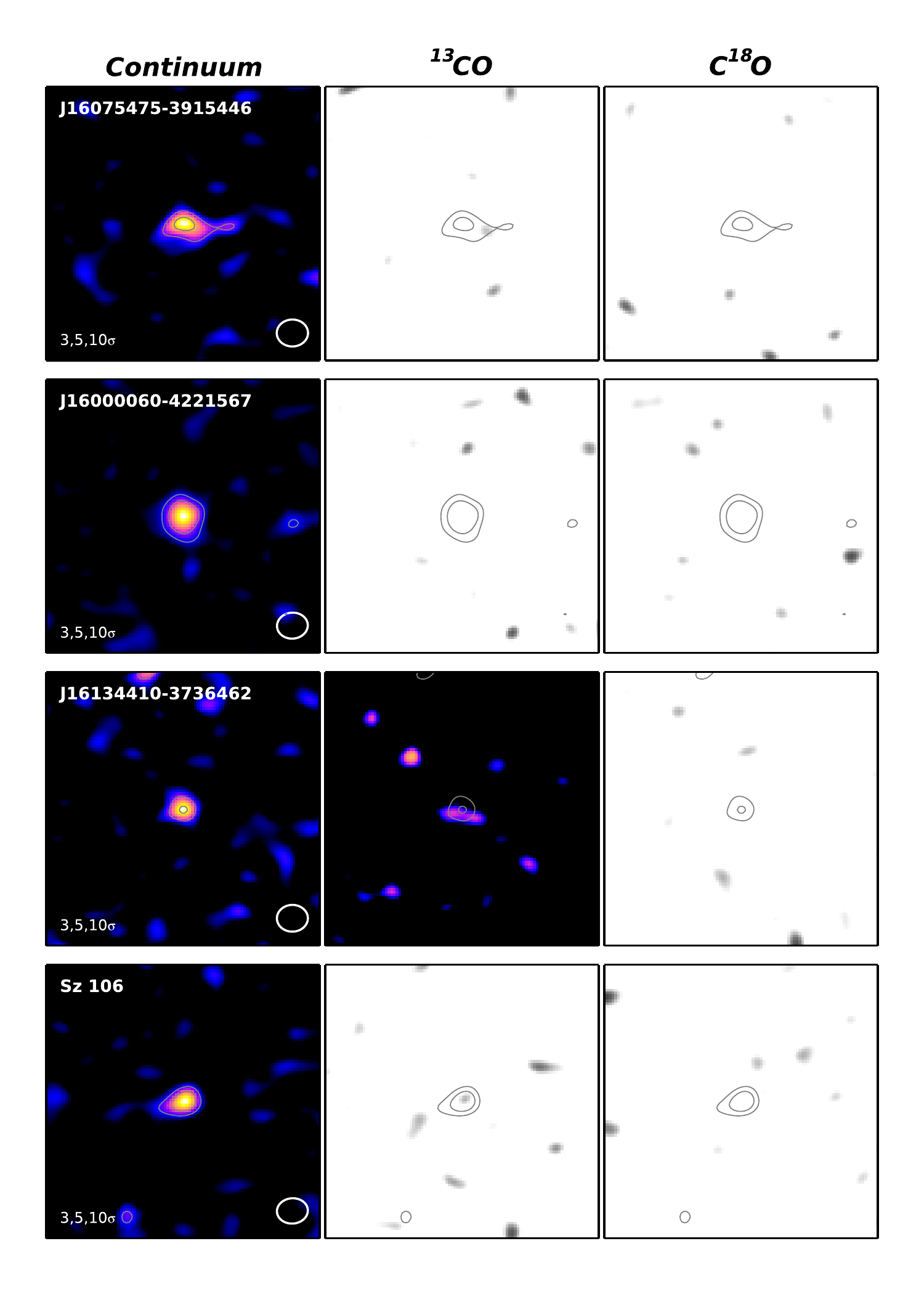}
\captcont{Cont.}
\end{center}
\end{figure*}

\begin{figure*}[!ht]
\begin{center}
\includegraphics[width=17cm]{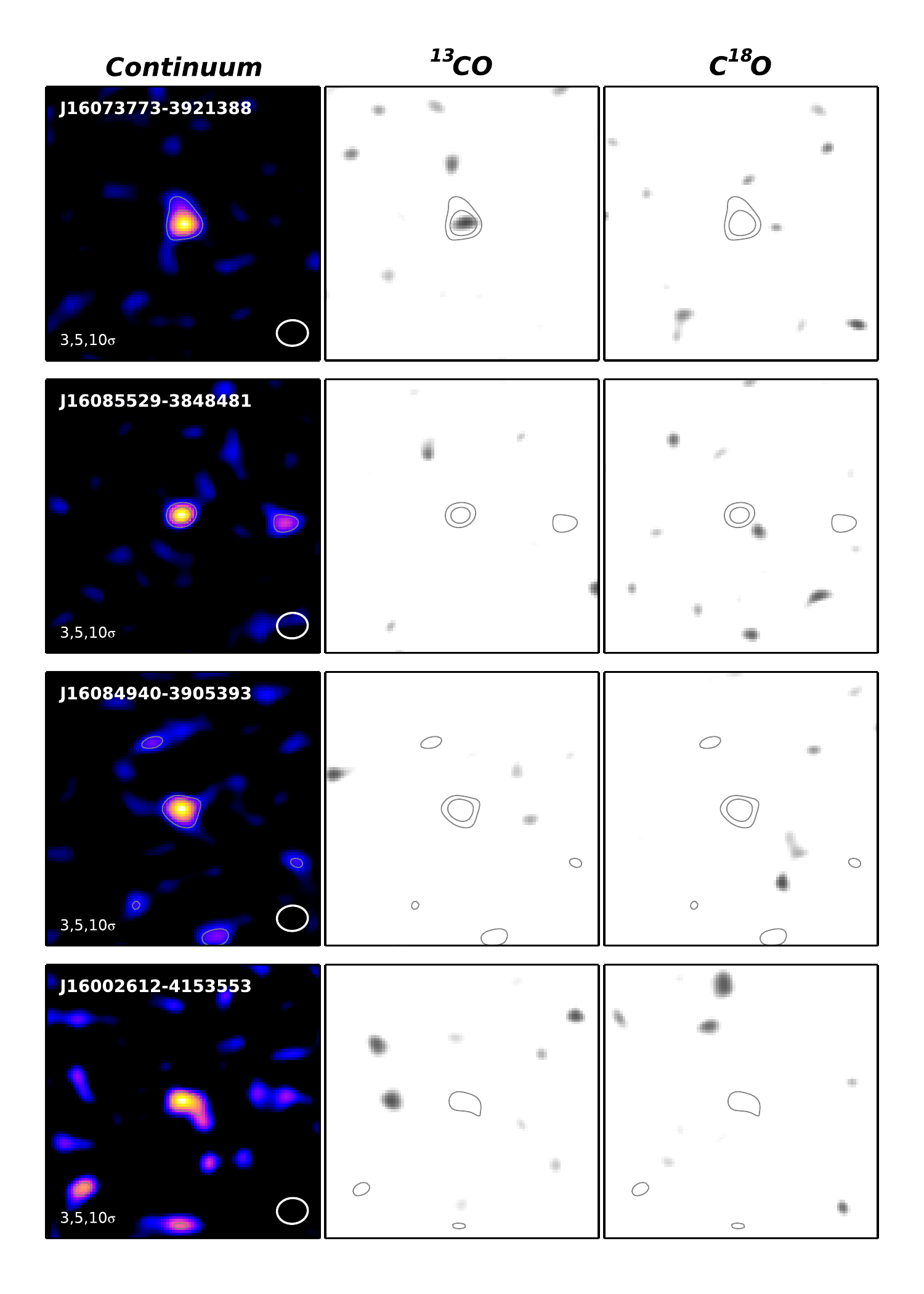}
\captcont{Cont.}
\end{center}
\end{figure*}

\begin{figure*}[!ht]
\begin{center}
\includegraphics[width=17cm]{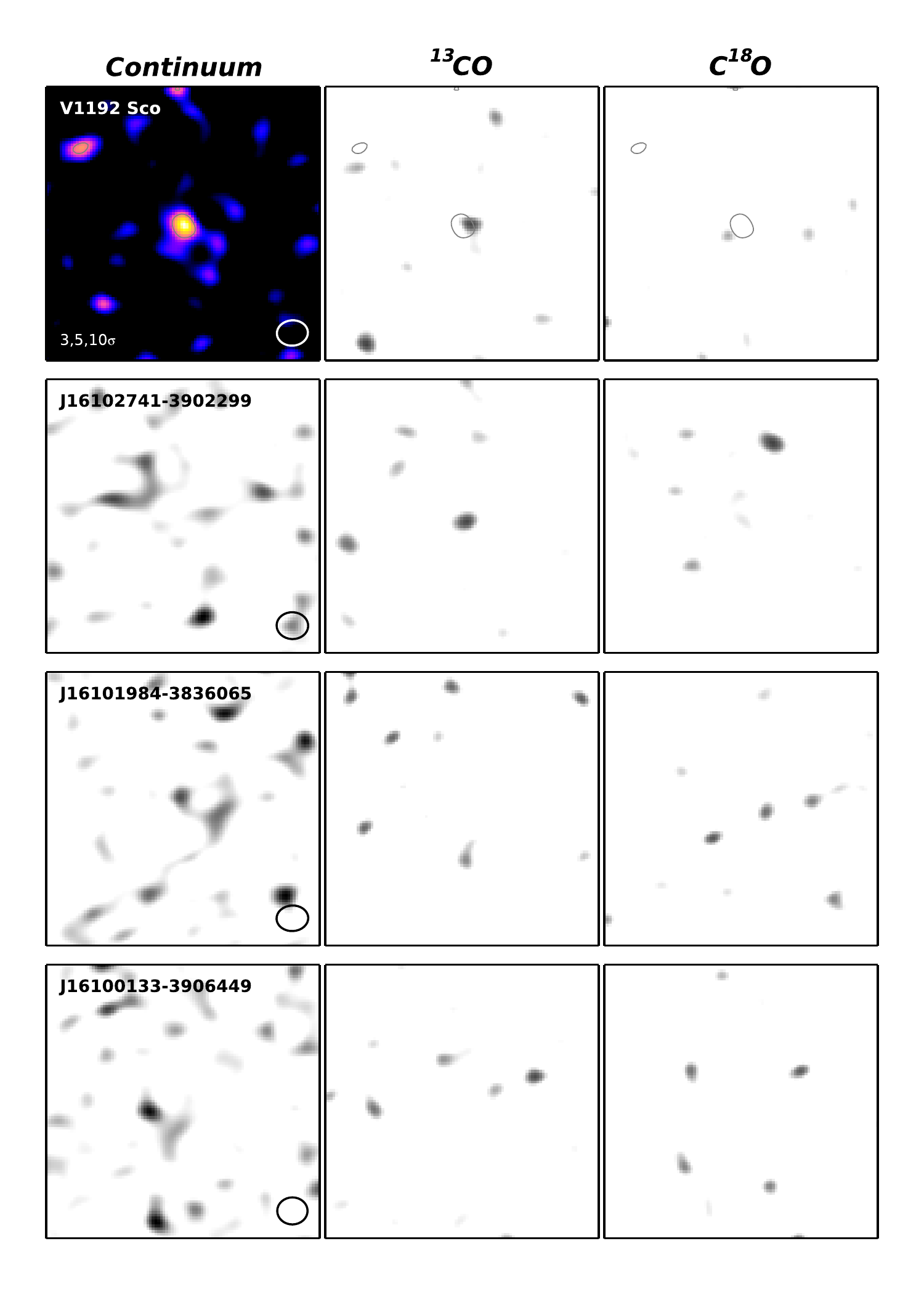}
\captcont{Cont.}
\end{center}
\end{figure*}

\begin{figure*}[!ht]
\begin{center}
\includegraphics[width=17cm]{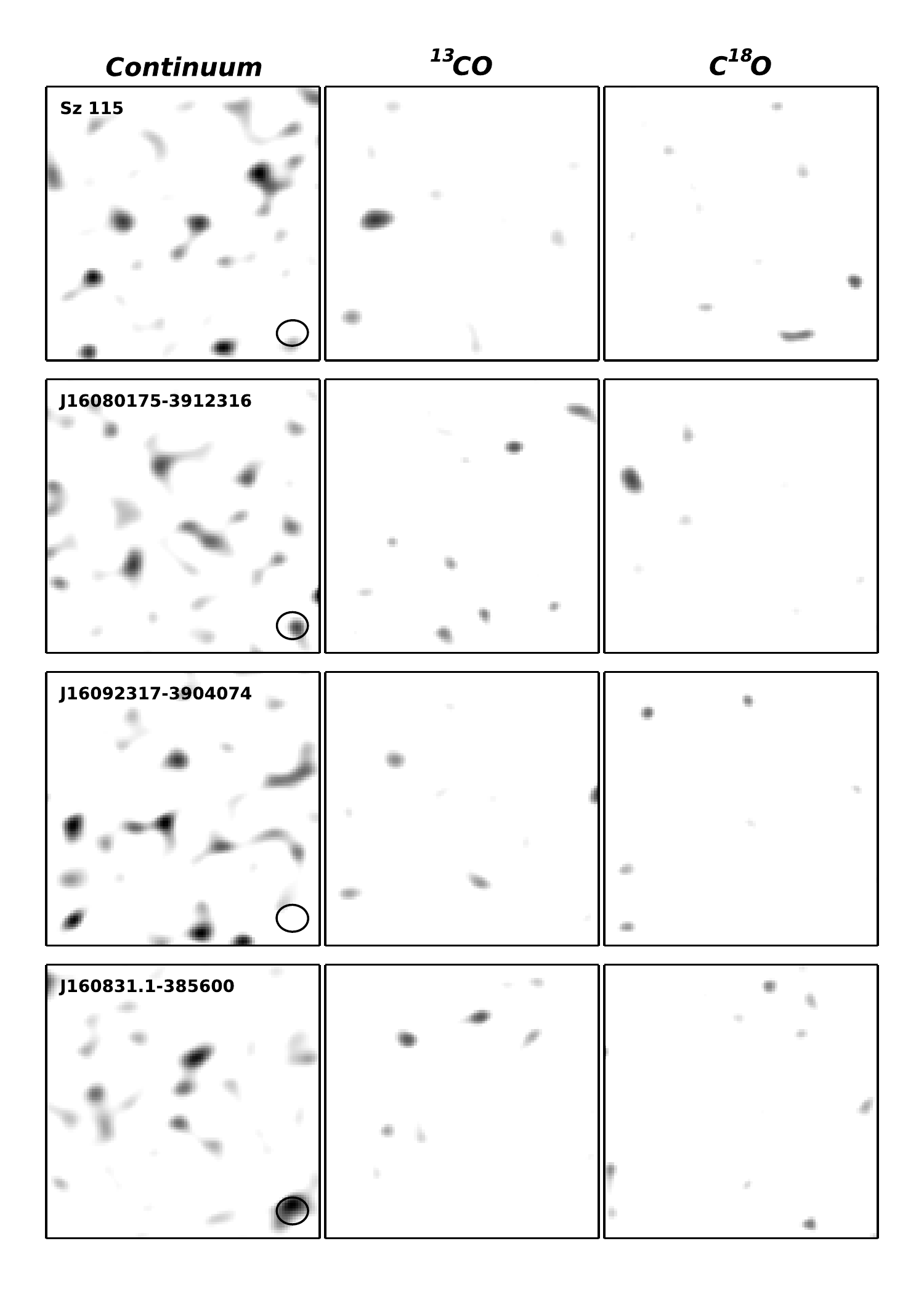}
\captcont{Cont.}
\end{center}
\end{figure*}

\begin{figure*}[!ht]
\begin{center}
\includegraphics[width=17cm]{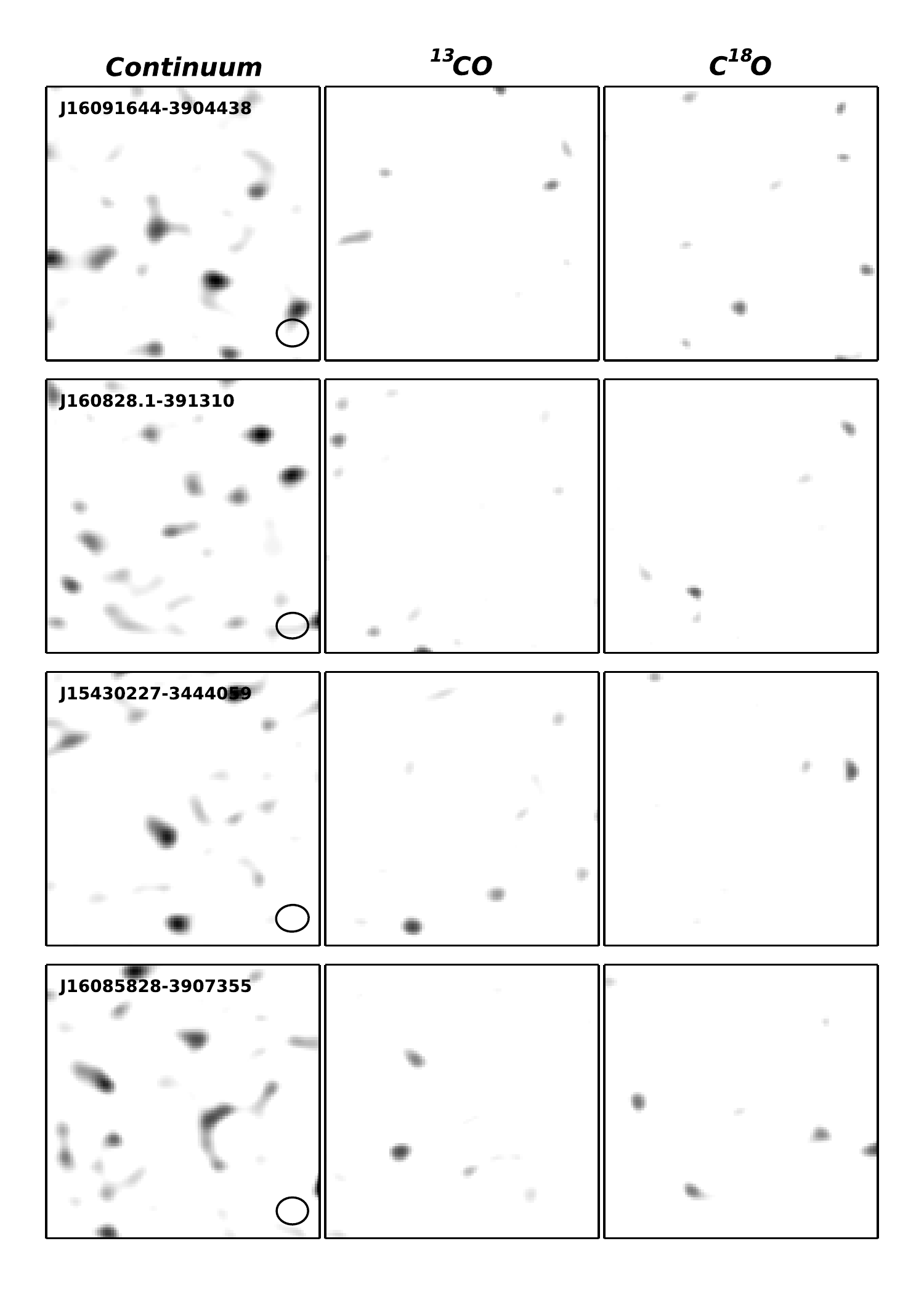}
\captcont{Cont.}
\end{center}
\end{figure*}

\begin{figure*}[!ht]
\begin{center}
\includegraphics[width=17cm]{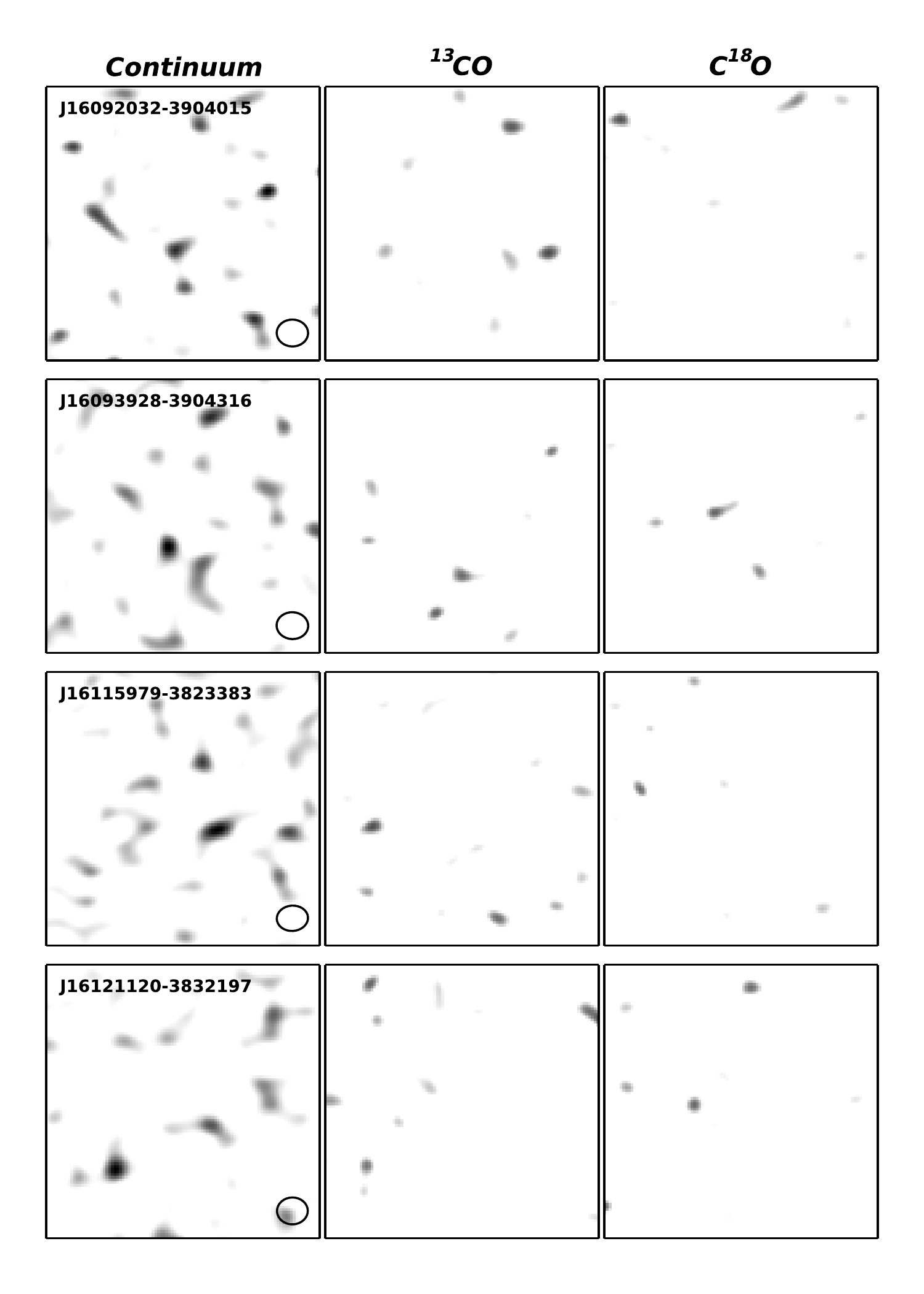}
\captcont{Cont.}
\end{center}
\end{figure*}

\begin{figure*}[!ht]
\begin{center}
\includegraphics[width=17cm]{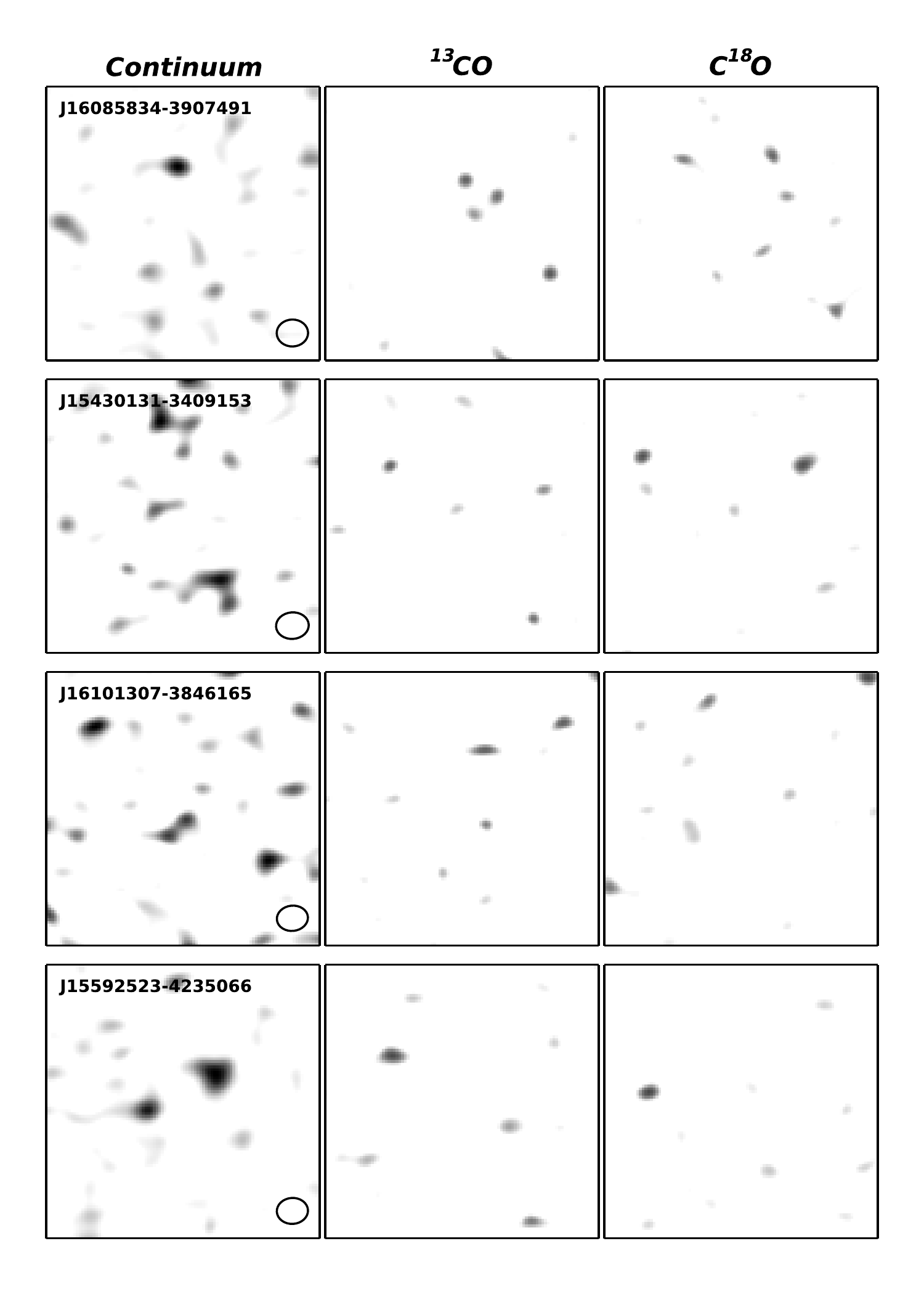}
\captcont{Cont.}
\end{center}
\end{figure*}

\begin{figure*}[!ht]
\begin{center}
\includegraphics[width=17cm]{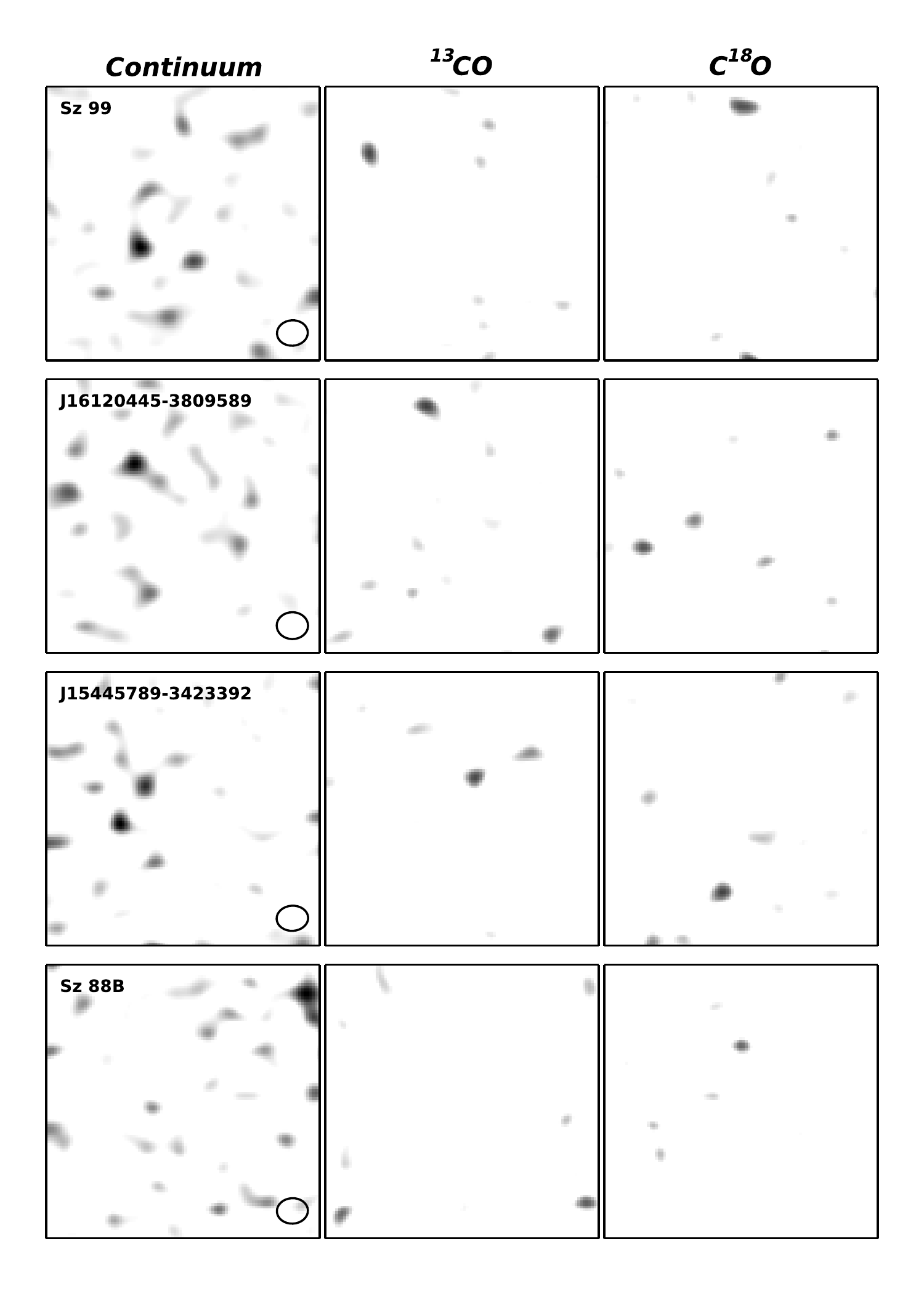}
\captcont{Cont.}
\end{center}
\end{figure*}

\begin{figure*}[!ht]
\begin{center}
\includegraphics[width=17cm]{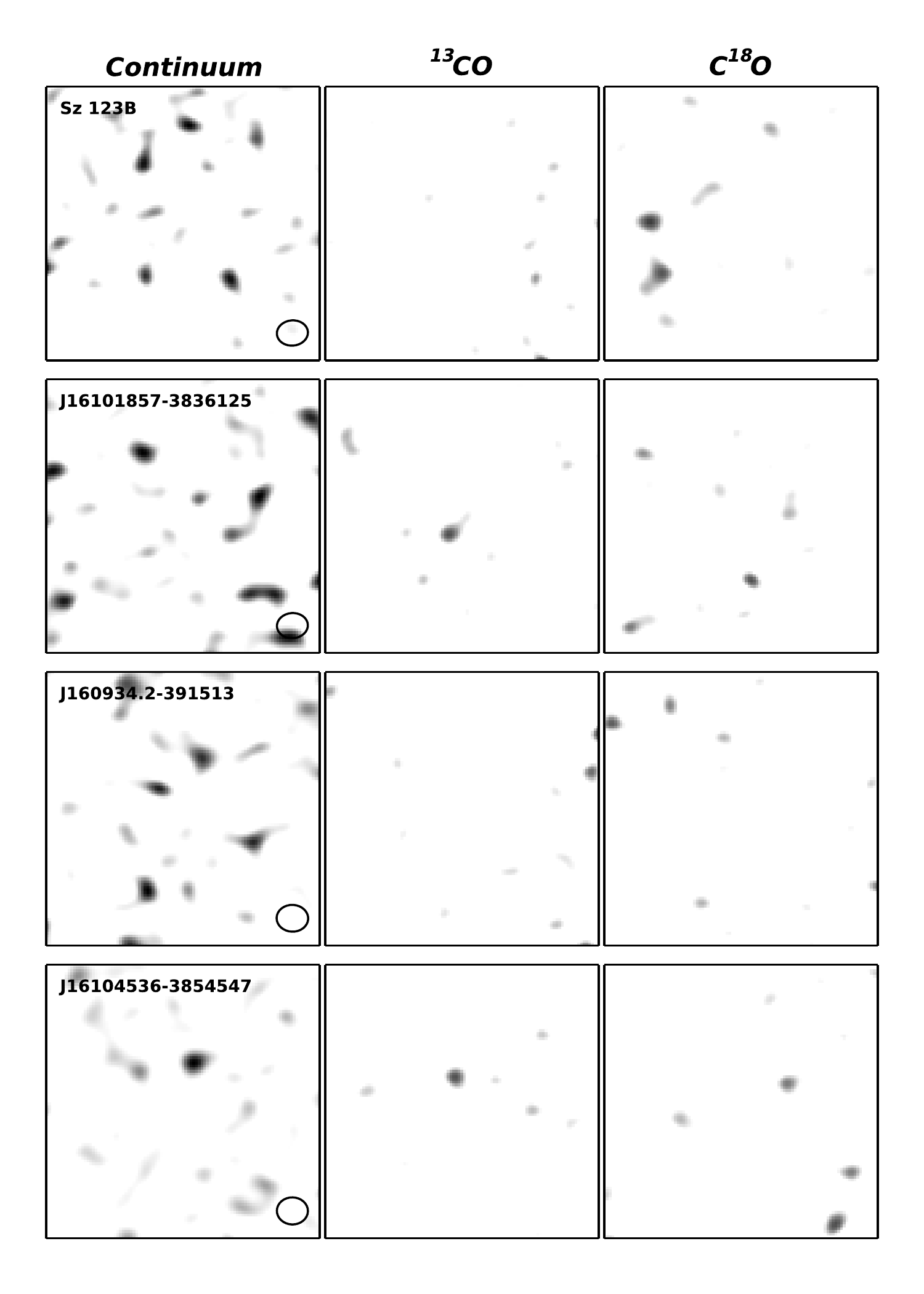}
\captcont{Cont.}
\end{center}
\end{figure*}

%\fi

\end{document}